\documentclass[twocolumn,prl,floatfix,preprintnumbers,a4paper,nofootinbib,superscriptaddress]{revtex4-2}
\usepackage[utf8]{inputenc}
\usepackage[T1]{fontenc}
\usepackage{amsmath,amssymb}
\usepackage{amsfonts}
\usepackage{bm}
\usepackage{courier}
\usepackage{graphics}
\usepackage{graphicx}
\usepackage{braket}
\usepackage{comment}
\usepackage{float}
\usepackage{amsmath}
\usepackage{amssymb}
\usepackage[normalem]{ulem}
\usepackage[mathscr]{euscript}
\usepackage{acronym}
\usepackage{float}
\usepackage[caption=false]{subfig}
\usepackage{multirow}
\usepackage{graphicx}
\usepackage{multirow}
\usepackage{graphicx}
\usepackage[usenames,dvipsnames,table,xcdraw]{xcolor}
\usepackage[colorlinks, pdfborder={0 0 0}]{hyperref} 
\usepackage[nameinlink,capitalise]{cleveref}
\usepackage{tensor}
\usepackage{verbatim}
\usepackage{wrapfig}
\usepackage{soul}

\graphicspath{{fig}}

\newcommand{\sapienza}{Dipartimento di Fisica, Sapienza Università 
	di Roma, Piazzale Aldo Moro 5, 00185, Roma, Italy}
\newcommand{\infn}{INFN, Sezione di Roma, Piazzale Aldo Moro 2, 00185, Roma, Italy}
\newcommand{\jhu}{William H. Miller III Department of Physics and Astronomy, Johns Hopkins University, 3400 North Charles Street, Baltimore, Maryland, 21218, USA}

\begin{document}
\title{Novel Ringdown Tests of General Relativity with Black Hole Graybody Factors}

\author{Romeo Felice Rosato}
\email{romeofelice.rosato@uniroma1.it}
\affiliation{\sapienza}
\affiliation{\infn}

\author{Francesco Crescimbeni}
\email{francesco.crescimbeni@uniroma1.it}
\affiliation{\sapienza}
\affiliation{\infn}

\author{Sophia Yi}
\email{syi24@jh.edu}
\affiliation{\jhu}

\author{Emanuele Berti}
\email{berti@jhu.edu}
\affiliation{\jhu}

\author{Paolo Pani}
\email{paolo.pani@uniroma1.it}
\affiliation{\sapienza}
\affiliation{\infn}

\begin{abstract}
We present \textsc{GreyRing}, a new model for the postmerger signal in black-hole binary coalescences based on the graybody factor of the remnant. 
The model accurately reproduces the full frequency-domain ringdown signal for a large set of comparable-mass, aligned-spin numerical relativity waveforms, including the $(\ell,m)=$(2,2), (3,3) and (4,4) modes. It achieves mismatches of order ${\cal O}(10^{-6})$ for the dominant $(2,2)$ mode, demonstrating that the graybody-factor parametrization provides a very accurate phenomenological description of the postmerger signal.
Building on this model, we introduce a novel consistency  test of strong gravity based on the graybody factor: the remnant mass and spin inferred from \textsc{GreyRing} can be compared with those obtained through standard black-hole spectroscopy. 
This agnostic test relies exclusively on the postmerger signal and does not require the inclusion of overtones or the choice of very early ringdown starting times, combining the advantages of inspiral-merger-ringdown consistency tests and traditional black-hole spectroscopy.
We apply the test to GW250114 and find that the remnant mass and spin inferred from \textsc{GreyRing} are consistent with those measured from the full signal. 
Remarkably, the inferred parameters can be measured with a precision 
comparable to, or slightly better than, that achieved with standard black-hole spectroscopy.
Our graybody-factor waveform model allows for new precision tests of strong gravity using the ringdown signal.
\end{abstract}

\maketitle
%%%%%%%%%%%%%%%%%%%%%%%%

%%%%%%%%%%%%%%%%%%%%%%%%%%%%
\noindent
{\bf \em Introduction.}
%%%%%%%%%%%%%%%%%%%%%%%%%%%%
The final stage of a binary black hole~(BH) merger, the so-called ``ringdown,'' provides one of the cleanest probes of strong-field gravity. 
During this phase the newly formed BH relaxes toward equilibrium through damped oscillations, and the gravitational waveform is commonly described as a superposition of quasinormal modes (QNMs) of the remnant spacetime~\cite{Berti:2025hly}. 
In general relativity (GR), the remnant is a rotating (Kerr) BH and the entire QNM spectrum is uniquely determined by its mass $M$ and  dimensionless spin $\chi=J/M^2$~\cite{Vishveshwara:1970zz,Press:1971wr,Teukolsky:1973ha,Chandrasekhar:1975zza,Kokkotas:1999bd,Berti:2009kk} (we use geometrical units, $G=c=1$). 
This property underlies the program of BH spectroscopy, in which multiple QNMs extracted from the data are checked for consistency with the Kerr prediction~\cite{Dreyer:2003bv,Berti:2005ys,Gossan:2011ha}. 
Such measurements enable stringent tests of GR~\cite{Berti:2015itd,Berti:2018vdi,Franchini:2023eda,LIGOScientific:2026qni,LIGOScientific:2026wpt}, of the nature of compact remnants~\cite{Cardoso:2019rvt,Maggio:2020jml,Maggio:2021ans,Maggio:2023fwy}, and of near-horizon environments~\cite{Barausse:2014tra,Cardoso:2021wlq}.

In practice, however, accurate ringdown measurements remain challenging. While BH perturbation theory predicts an infinite tower of QNMs, only a few are expected to be observable in realistic mergers. 
Even in the loudest event observed to date, GW250114, only two QNMs have been confidently measured, allowing for a ringdown test of GR at the level of $\approx30\%$ for the frequency of the prograde quadrupolar first overtone
~\cite{LIGOScientific:2025rid,LIGOScientific:2025obp}.
% prograde (2,2,1) mode} 
% 
%
The starting time of the ringdown is inherently uncertain and the number of modes that should be included in the analysis must be carefully guided by theoretical models. Only certain QNM sets are excited and detectable at any given time postmerger~\cite{Crescimbeni:2025ytx}, and ringdown tests that include too many free parameters can overfit the data. This has fueled an ongoing debate on the robustness of ringdown measurements (see e.g.~\cite{Isi:2019aib,Cotesta:2022pci,Isi:2023nif,Carullo:2023gtf}, and~\cite{Berti:2025hly} for a review).

Because only a small subset of QNMs can be measured, an independent determination of $M$ and $\chi$ is highly desirable. 
Current consistency tests estimate $M$ and $\chi$ either from the inspiral-merger signal or from the full waveform~\cite{LIGOScientific:2020tif,LIGOScientific:2021sio,LIGOScientific:2026wpt}, and compare these values with those inferred from the ringdown. 
In this approach, systematics in the modeling of the preringdown signal may
propagate into the inspiral-merger or full inspiral-merger-ringdown estimate of the remnant
properties~\cite{Gupta:2024gun,Dhani:2024jja} (see~\cite{CalderonBustillo:2020rmh,Chandra:2025ipu,Gennari:2023gmx,Kankani:2025gqj}
for proposed mitigation strategies based on numerical relativity surrogates and postpeak analyses).
For instance, unmodeled eccentricity or precession effects in the inspiral-merger analysis  could bias the inferred values of $M$ and $\chi$, potentially mimicking a deviation from GR~\cite{Narayan:2023vhm,Gupta:2024gun}.
These considerations motivate the search for complementary descriptions of the postmerger signal that do not rely exclusively on a QNM decomposition.

In this context, recent attention has focused on the BH graybody factor, which characterizes the scattering of perturbations by the spacetime curvature surrounding the BH. 
The graybody factor depends only on the underlying geometry, so it is a natural ``no-hair'' quantity that can be used to infer $M$ and $\chi$ from the observed ringdown waveform. 
Moreover, graybody factors are stable under small perturbations of the effective potential~\cite{Rosato:2024arw,Oshita:2024fzf,Rosato:2025lxb}, in contrast with the well-known spectral instability of QNMs~\cite{Nollert:1996rf,Barausse:2014tra,Jaramillo:2020tuu,Cheung:2021bol,Berti:2022xfj}.
Recent work suggested that the graybody factor modulates the ringdown amplitude in the frequency domain~\cite{Oshita:2023cjz}. 
Although originally investigated within BH perturbation theory for a plunging test particle~\cite{Oshita:2023cjz,Rosato:2024arw,Rosato:2025byu}, this description was later shown to capture remarkably well the spectral amplitude of comparable-mass mergers~\cite{Okabayashi:2024qbz,Rosato:2025ulx}. 

In this Letter, we present \textsc{GreyRing}, a new model for the postmerger ringdown signal based on the graybody factor of the remnant. 
Building on recent theoretical insights~\cite{Rosato:2026moe}, the model accurately reproduces {\em both} the amplitude and the phase of the frequency-domain postmerger signal, extending previous models that were limited to the amplitude~\cite{Oshita:2023cjz,Rosato:2024arw,Rosato:2025byu,Rosato:2025ulx}. 
We show that \textsc{GreyRing} reproduces a large representative set of comparable-mass, aligned-spin numerical relativity waveforms with typical mismatches of order ${\cal O}(10^{-6})$ for the dominant
quadrupole
% $\ell=m=2$
mode, thus enabling new ringdown tests of gravity with current and future detectors.

%%%%%%%%%%%%%%%%%%%%%%%%%%%%
\noindent
\textbf{\textsc{GreyRing}.}
%%%%%%%%%%%%%%%%%%%%%%%%%%%%
The gravitational-wave signal in the time domain can be written as
\begin{equation}
h_{+}-ih_{\times}= {M_T \over d_L}\sum_{\ell m} h_{\ell m} (t) \,{}_{-2}Y_{\ell m}(\theta,\phi)\,,
\end{equation}
where $h_{+}$ and $h_{\times}$ denote the plus and cross polarizations of the gravitational wave, $M_T$ is the total mass of the system, $d_{L}$ is the luminosity distance to the source, and  ${}_{-2}Y_{\ell m}$ are spin-weighted spherical harmonics with angular multipole numbers $(\ell,m)$, evaluated at the polar angle $\theta$ (inclination) and azimuthal angle $\phi$ specifying the direction of the observer in the source frame.
The frequency-domain counterpart of each multipole is $\tilde{h}_{\ell m}(\omega)=\int_{-\infty}^{+\infty} dt\, e^{i \omega t} h_{\ell m}(t)$.
Recently, Ref.~\cite{Rosato:2026moe} showed that the graybody factor is directly encoded in the component of the Green's function governing the linear ringdown response of a BH in the frequency domain, providing an analytical foundation for previous phenomenological proposals describing the spectral amplitude of the signal~\cite{Oshita:2023cjz,Rosato:2024arw,Rosato:2025byu,Rosato:2025ulx}.
Based on this theoretical insight, we propose to model each multipole as
\begin{equation}\label{eq:fullmodel}
\tilde h_{\ell m} (\omega)= {MA_{\ell m} \over (M\omega)^{p_{\ell m}}}\,R_{\ell m}(M\omega, \chi)\,e^{i {c_{\ell m} \over M\omega}} e^{i(\phi_c + \omega t_c)}\,,
\end{equation} 
where $M$ and $\chi$ are the remnant BH's mass and spin.
Here, $R_{\ell m}$ is the complex reflection amplitude, defined in terms of the graybody factor as the ratio between the outgoing and ingoing wave amplitudes in the remnant geometry (see Supplemental Material~\cite{SupplementalMaterial} and~\cite{Teukolsky:1972my,
Pani:2013pma,
Press:1972zz,
Press:1973zz,
Teukolsky:1974yv,
Starobinskii:1973vzb,
Starobinskil:1974nkd,
Brito:2015oca}); $\phi_c$ and $t_c$ are the phase and time of reference, respectively; and $A_{\ell m}$, $p_{\ell m}$, and $c_{\ell m}$ are source-dependent fitting parameters. 
In particular, the model for the amplitude,
\begin{equation}\label{eq:amplitude}
|\tilde h_{\ell m} (\omega)|= M\,{A_{\ell m} \over (M\omega)^{p_{\ell m}}}\,|R_{\ell m}(M\omega, \chi)|\,,
\end{equation}
was previously validated against numerical relativity simulations of binary BH mergers, and $(A_{\ell m},p_{\ell m})$ were fitted as functions of the progenitor parameters~\cite{Rosato:2025ulx}. Here, for the first time, we complete the waveform template by including a model for the phase
%, while
%
\begin{equation}\label{eq:phase}
\arg\!\left(\tilde h_{\ell m} (\omega)\right)=\arg\!\left(R_{\ell m}(M\omega, \chi)\right)+\omega t_c+\phi_c+ {c_{\ell m}\over M\omega}\,.
\end{equation}
The term $\sim c_{\ell m}/(M\omega)$ is the simplest among a family of parametrizations discussed in Supplemental Material~\cite{SupplementalMaterial}. It can be viewed as the leading term in a high-frequency expansion of the residual source-induced dephasing, after the trivial time and phase shifts have been factored out.

Overall, each $(\ell,\,m)$ multipole is modeled by seven real parameters: $(M, \chi,\phi_c,\,t_c)$, and three source-dependent parameters ($A_{\ell m}$,\,$p_{\ell m}$,\,$c_{\ell m}$).
\begin{figure}[t]
    \centering
    \includegraphics[width=\linewidth]{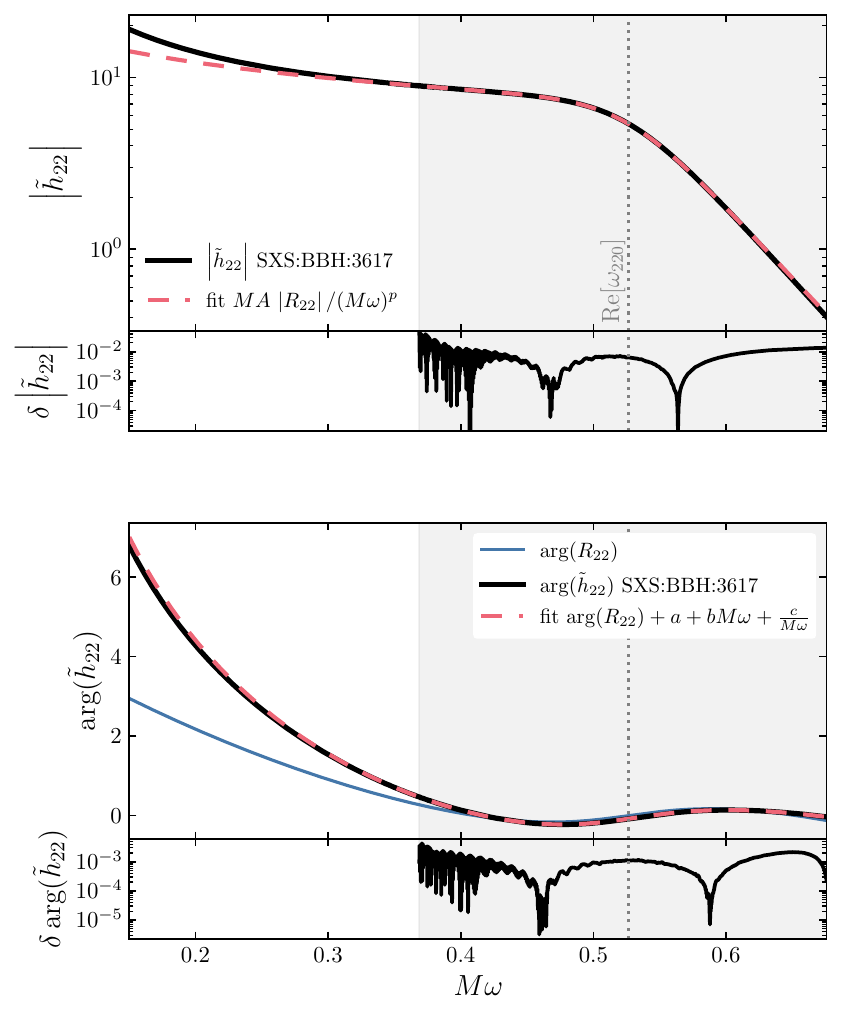}
    \caption{
    Fit of the \textsc{GreyRing} model to the postmerger quadrupolar signal of a comparable-mass binary BH simulation (\texttt{SXS:BBH:3617}). 
    Top panel: amplitude of the Fourier-domain waveform (black), compared with the model in Eq.~\eqref{eq:amplitude} (dashed orange). 
    Bottom panel: detrended phase of the numerical signal (black) and of the graybody factor $R_{22}$ (blue), together with the fitted model including the $c_{\ell m}/(M\omega)$ correction (dashed orange); see Eq.~\eqref{eq:phase}. 
    The dashed orange curves are also extrapolated outside the fitting window (gray shaded region) toward lower frequencies, where they still provide a reasonable description of the numerical signal, especially for the phase.
    As a reference, the dotted vertical line marks the real part of the fundamental QNM. In the bottom inset of each panel, we plot the residual $\delta X =|X_{\rm SXS}-X_{\rm fit}|$ for the corresponding quantity $X$.
    }
    \label{fig:fit}
\end{figure}

\noindent
{\bf \em Validation against numerical simulations.}
We validate the \textsc{GreyRing} model against $311$ comparable-mass, quasicircular, nonprecessing binary BH simulations from the Simulating eXtreme Spacetimes~(SXS) Collaboration catalog~\cite{Boyle:2019kee,Scheel:2025jct}, with mass ratios $q\in[1,4]$. We perform two complementary tests.

First, we assess whether the model can recover the properties of the remnant geometry. To this end, we employ the mismatch
\begin{equation}\label{eq:mismatch} \mathcal{M}= 1- \frac{\braket{h^{\rm NR}_{\ell m} |h^{\rm model}_{\ell m}}} {\sqrt{\braket{h^{\rm NR}_{\ell m}|h^{\rm NR}_{\ell m}} \braket{h^{\rm model}_{\ell m}|h^{\rm model}_{\ell m}}}}\,, \end{equation} 
where $h^{\rm NR}_{\ell m}$ is the numerical relativity waveform, $h^{\rm model}_{\ell m}$ is given by Eq.~\eqref{eq:fullmodel}, and the inner product is defined without detector-noise weighting, i.e.,  \begin{equation} 
\braket{h_1|h_2} = \int_{\omega_i}^{\omega_f} d\omega \, \tilde h_1(\omega)\,\tilde h_2^*(\omega)\,, 
\end{equation} 
as our goal is to assess the model's intrinsic accuracy. We minimize $\mathcal{M}$ by varying the seven model parameters. The best-fit values $(M_{\rm fit},\chi_{\rm fit})$ are then compared with the remnant quantities reported in the SXS metadata. Across the catalog, \textsc{GreyRing} recovers the remnant mass and spin with typical relative errors of order $10^{-3}$ for the dominant $(2,2)$ mode, both in $|M_{\rm fit}-M_{\rm NR}|/M_{\rm NR}$ and in $|\chi_{\rm fit}-\chi_{\rm NR}|$, consistent with the accuracy obtained in standard QNM-based remnant-parameter recovery~\cite{Giesler:2019uxc,Baibhav:2023clw}.
Details of the analysis are presented in Supplemental Material~\cite{SupplementalMaterial}.

Second, fixing $M$ and $\chi$ to their numerical relativity values, we quantify the intrinsic waveform accuracy of the model through the mismatch between numerical and modeled multipoles.
In Fig.~\ref{fig:fit} we illustrate the performance of the model on the numerical relativity simulation \texttt{SXS:BBH:3617}. The model is fitted over the frequency interval $\omega \in [\omega_i,\omega_f]$, chosen following the same stability criterion introduced in Ref.~\cite{Rosato:2025ulx} and here applied consistently to the extended \textsc{GreyRing} model.
In the upper panel we show that the model of Eq.~\eqref{eq:amplitude} accurately reproduces the numerical amplitude $|\tilde{h}_{\ell m}(\omega)|$. In the lower panel we present a similar comparison for the phase. We detrend the numerical phase by subtracting its best-fit linear component, related to the term $\omega t_c + \phi_c$ in Eq.~\eqref{eq:phase}. A minimal model in which we similarly detrend  the phase of $R_{\ell m}(M\omega,\chi)$ (shown in blue) already exhibits very good agreement. 
Including the subleading term $c_{\ell m}/(M\omega)$ yields excellent agreement between the numerical phase and the model, shown as a dashed orange line, across the entire frequency range.

We do not extremize over $t_c$ and $\phi_c$, as these degrees of freedom are already included within the phase model and determined by the fitting procedure.
For the case shown in Fig.~\ref{fig:fit}, we find $\mathcal{M} \simeq 2.0 \times 10^{-6}$.

%%%%
\begin{figure}[t]
\centering
\includegraphics[width=\linewidth]{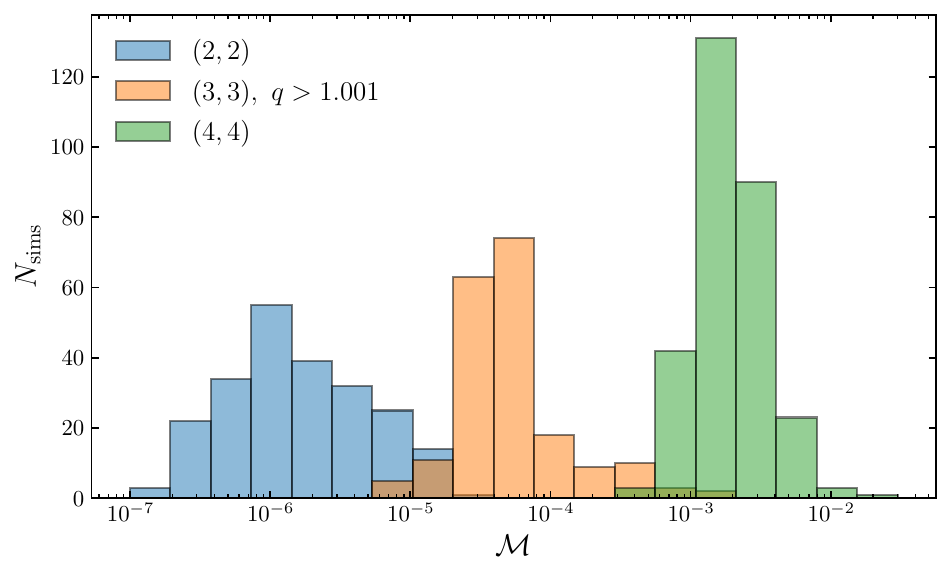}
\caption{
Mismatch distribution of the \textsc{GreyRing} model for the $(\ell,\,m)=(2,\,2),\,(3,\,3)$ and $(4,\,4)$ multipoles over $311$ comparable-mass, spin-aligned binary BH simulations from the SXS catalog~\cite{Boyle:2019kee,Scheel:2025jct}. For the $(3,\,3)$ mode, we restrict to the subset of simulations with mass ratio $q>1.001$, because the mode amplitude vanishes in the equal-mass limit. }
\label{fig:testmodel}
\end{figure}
%%%%

We test the model against $311$ SXS simulations of comparable-mass, quasicircular, nonprecessing binary BHs with mass ratios in the range $[1,4]$~\cite{Boyle:2019kee,Scheel:2025jct}.
In Fig.~\ref{fig:testmodel} we show the mismatch distribution across the catalog for the $\ell=m=2,\,3,\,4$  multipoles.
For $(\ell,m)=(2,\,2)$, \textsc{GreyRing} achieves mismatches in the range ${\cal M}\in [3\times 10^{-8},5\times 10^{-5}]$, and typically ${\cal M}={\cal O}(10^{-6})$.
The mismatch for $(\ell,m)=(3,\,3)$ and $(\ell,m)=(4,\,4)$ is at the level of ${\cal O}(10^{-5})$ and ${\cal O}(10^{-3})$, respectively.
While the reduced accuracy may partly reflect the numerical precision of the SXS catalog, we also expect that incorporating effects currently neglected in our model --~such as spherical-spheroidal mode mixing~\cite{Berti:2014fga} and nonlinear mode coupling~\cite{Cheung:2022rbm,Mitman:2022qdl,Cardoso:2026llh}~-- will further improve the agreement for higher-order multipoles.
Nevertheless, when the modeled multipoles are combined to reconstruct the full angular dependence of the waveform, the mismatch remains typically $\mathcal{O}(10^{-5})$ because the signal is dominated by the $(2,2)$ mode. A detailed discussion is provided in Supplemental Material~\cite{SupplementalMaterial}.

The mismatches for the present model are at least one order of magnitude lower than for the amplitude-only model~\cite{Rosato:2025ulx}. Remarkably, typical \textsc{GreyRing} model mismatches are comparable to (or better than) state-of-the-art time-domain models calibrated using numerical relativity simulations~\cite{Crescimbeni:2025ytx}, such as the \texttt{TEOBPM} template~\cite{Damour:2014yha,DelPozzo:2017rka,Nagar:2020eul,Nagar:2020xsk}.
%

%%%%%%%%%%%%%%%%%%%%%%%%%%%%
\noindent
{\bf \em Tests of gravity with \textsc{GreyRing}.} 
%%%%%%%%%%%%%%%%%%%%%%%%%%%%
The high accuracy of our model enables a novel agnostic test of GR: we can use \textsc{GreyRing} to infer $M$ and $\chi$ from the frequency-domain signal, and compare the resulting posteriors with those obtained with a standard QNM-based ringdown template. This is conceptually similar to inspiral–merger–ringdown consistency tests, which infer the remnant properties from the progenitor binary using fits calibrated to numerical relativity simulations in GR~\cite{LIGOScientific:2026qni}. Unlike those tests, our approach relies exclusively on the postmerger signal, and it does not require connecting the inspiral to the ringdown via numerical relativity modeling.
Moreover, a consistency test can be performed by extracting $(M,\,\chi)$ using only the $\ell=m=2$ multipole in \textsc{GreyRing} and only the fundamental $\ell=m=2$ QNM in standard BH spectroscopy, without the need for higher modes~\cite{Gossan:2011ha,Brito:2018rfr} or overtones~\cite{Isi:2019aib,Giesler:2019uxc,Ota:2019bzl,Bhagwat:2019dtm,CalderonBustillo:2020rmh,Cotesta:2022pci,Isi:2022mhy,Finch:2022ynt,Carullo:2023gtf,Ma:2023vvr,Ma:2023cwe,Baibhav:2023clw}.
This is a significant advantage, because (i) $\ell>2$ modes are generically harder to measure; (ii) for highly symmetric binaries -- including GW250114~\cite{LIGOScientific:2025epi,LIGOScientific:2025obp}, which has enabled the most stringent ringdown tests to date~\cite{LIGOScientific:2026wpt} --  odd-$m$ multipoles are strongly suppressed; and (iii) multimode QNM fits are subject to ambiguities in the starting time and in the number of basis elements retained in the template. These become especially relevant when overtones are included, since the nonorthogonality of the fitted basis can lead to overfitting~\cite{JimenezForteza:2020cve,Cotesta:2022pci,Baibhav:2023clw}.

As a first step, we have applied this test by performing Bayesian parameter estimation on synthetic data (see Supplemental Material~\cite{SupplementalMaterial} for details). 
We assume nonprecessing binaries, so that $\pm m$ multipoles are related by $\tilde h_{\ell m}(\omega)=(-1)^\ell \tilde h^*_{\ell\,-m}(-\omega)$.
We consider the GW250114-like numerical simulation \texttt{SXS:BBH:3617} and perform injection–recovery tests with two independent models. We first generate injections using the QNM model calibrated to the simulation and perform a standard QNM analysis with \texttt{pycbc inference}~\cite{Biwer:2018osg}. We inject QNMs with physical parameters consistent with GW250114 in the $\ell=m=2$ multipole and recover the posterior distributions of $(M,\,\chi)$. 

Then, we perform an independent parameter estimation using \textsc{GreyRing}, in which we inject a signal constructed from the same simulation and recover it using only the frequency-domain $\ell=m=2$ multipole.
As expected for consistency, the posteriors on $(M,\,\chi)$ recovered in this case are compatible with those from BH spectroscopy. Furthermore, the uncertainties of the two methods are similar.

%%%%%%%%%%%%%%%%%%%%%%%%%%%%
\noindent
{\bf \em  Application to GW250114.}
%%%%%%%%%%%%%%%%%%%%%%%%%%%%
We apply the \textsc{GreyRing} consistency test to real data, focusing on GW250114. For this event (the loudest detected so far) the real and imaginary parts of the $\ell=m=2$ fundamental QNM frequency were measured at the level of $6\%$ and $10\%$, the first $\ell=m=2$ overtone frequency was measured with an accuracy of approximately $30\%$, and there is some evidence for other modes in the ringdown~\cite{LIGOScientific:2025epi,LIGOScientific:2025obp}.

\begin{figure}[t]
    \centering
    \includegraphics[width=0.97\linewidth]{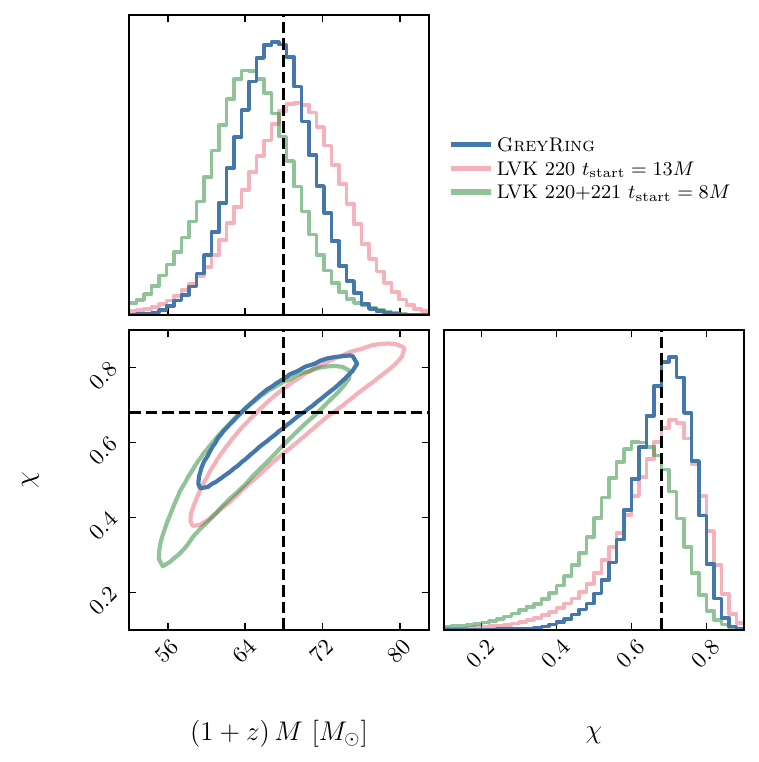}
    \caption{
    $90\%$ credible region of the joint posterior distribution for the redshifted remnant mass and spin of GW250114. The blue contour corresponds to our graybody-factor model applied in the frequency range $f\in[100,512]\,\rm{Hz}$. The red and green contours show the results of a standard QNM analysis performed with \texttt{pyRing}~\cite{pyRing} using the $220$ mode and the $220+221$ modes, respectively~\cite{LIGOScientific:2025epi,LIGOScientific:2025obp,ligo_virgo_kagra_2025_17018009}. Black dashed lines mark the maximum-likelihood estimate from the LIGO-Virgo-KAGRA analysis with the \texttt{NRSur7dq4} inspiral-merger-ringdown waveform model~\cite{Varma:2019csw}. 
    }
    \label{fig:GW250114}
\end{figure}

Our main results are shown in Fig.~\ref{fig:GW250114}, where we apply \textsc{GreyRing} to the Fourier transform of the GW250114 time series. The resulting posteriors (blue contours) are compared with the standard LIGO-Virgo-KAGRA ringdown analysis using either the fundamental $\ell=m=2$ mode alone (red) or a model including also the first overtone (green)~\cite{LIGOScientific:2025epi,LIGOScientific:2025obp, ligo_virgo_kagra_2025_17018009}. The ringdown analysis starts at $t_{\rm start}=13M$ or $t_{\rm start}=8M$ after the peak of the strain in the two cases, respectively, since including the first overtone better captures the early postmerger phase~\cite{Isi:2019aib,Giesler:2019uxc}. 
The \textsc{GreyRing} posteriors are compatible with (and slightly narrower than) those from the standard ringdown analysis~\cite{LIGOScientific:2025epi,LIGOScientific:2025obp}, yielding $(1+z)M= 66.9_{-6.4}^{+6.1}\,M_\odot $ and $\chi=0.69_{-0.16}^{+0.10}$ (90\% credible intervals) for the final detector-frame mass and spin, respectively.
The lower cutoff frequency in Fig.~\ref{fig:GW250114} is chosen using the stability criterion of Ref.~\cite{Rosato:2025ulx}, applied here to real-data posteriors: we select the range where the inferred remnant parameters are stable under changes in $f_{\min}$ and not yet affected by the loss of high-frequency signal power. The Supplemental Material~\cite{SupplementalMaterial} presents the full corner plots, an independent validation of the standard ringdown analysis, and variations of both $t_{\rm start}$ and the frequency range, showing that the \textsc{GreyRing} posteriors on $M$ and $\chi$ remain compatible and stable as $f_{\min}$ is varied, unlike the stronger starting-time dependence of standard QNM spectroscopy.

%%%%%%%%%%%%%%%%%%%%%%%%%%%%
\noindent
{\bf \em  Discussion.}
%%%%%%%%%%%%%%%%%%%%%%%%%%%%
Our graybody-factor model opens up a new class of tests of gravity in the ringdown regime. Although we have focused here on the dominant $\ell=m=2$ mode, analogous tests can be performed with subdominant harmonics, in parallel with standard QNM-based studies. 

Compared to ordinary BH spectroscopy, \textsc{GreyRing} offers several key advantages:

\begin{itemize}
\item[(i)] The graybody factor encodes the full linear response of the remnant in the frequency domain and does not require a QNM decomposition. This effectively resums \emph{all} overtones and captures features of the response that are not included in a QNM expansion~\cite{Andersson:1996cm,Leaver:1986gd,DeAmicis:2025xuh,DeAmicis:2026tus,Su:2026fvj,Aruquipa:2026tga,Rosato:2026moe}, such as late-time tails~\cite{Rosato:2026moe}.

\item[(ii)] Unlike QNMs, the graybody factor is a complex, frequency-dependent function. The consistency tests proposed here probe the \emph{entire} functional form of $R_{\ell m}(M\omega,\chi)$ rather than a discrete set of frequencies and damping times, thereby exploiting more of the signal's information to reconstruct a no-hair property of the remnant. 

\item[(iii)] Compared to standard QNM measurements, the graybody factor captures a broad frequency range, including relatively high frequencies. Therefore, it is expected to be more sensitive to short-length-scale modifications in the strong-field regime, such as higher-order curvature corrections to GR~\cite{Antoniou:2026jvh} or horizon-scale physics~\cite{Rosato:2025byu,Rosato:2025lxb}.
\end{itemize}

The \textsc{GreyRing} model can be extended to perform additional consistency tests of gravity that are complementary to (and independent of) traditional inspiral-merger-ringdown tests. For example, a less conservative implementation of the agnostic test proposed in this Letter would rely on numerical relativity simulations to map the progenitor masses and spins to the model parameters ($A_{\ell m}$,\,$p_{\ell m}$,\,$c_{\ell m}$) along the lines of Ref.~\cite{Rosato:2025ulx}, thus enabling consistency checks with the information extracted from the inspiral signal.

The increasing sensitivity of current detectors, together with the prospects of next-generation ground- and space-based observatories such as the Einstein Telescope~\cite{ET:2019dnz,Branchesi:2023mws,ET:2025xjr}, Cosmic Explorer~\cite{Reitze:2019iox,Evans:2021gyd} and LISA~\cite{LISA:2024hlh}, strongly motivates the development of accurate and robust frameworks to parametrize and constrain deviations from the Kerr metric in the ringdown regime. Systematic effects will be dominant at large signal to noise, particularly relevant in the context of ringdown, where modeling uncertainties in the starting time and in the number of modes can significantly bias the inference~\cite{Capuano:2025kkl,Volkel:2025jdx}. In this context, approaches that minimize such systematics are especially valuable~\cite{Gupta:2024gun,LISAConsortiumWaveformWorkingGroup:2023arg}.

\textsc{GreyRing} achieves mismatches of order ${\cal O}(10^{-6}-10^{-5})$ for signals including modes up to $\ell=m=4$, typically outperforming state-of-the-art time-domain models~\cite{Crescimbeni:2025ytx}.
Using the conservative criterion that two waveforms are indistinguishable for parameter-estimation purposes when $\mathcal{M} \lesssim (2\,{\rm SNR}^2)^{-1}$~\cite{Flanagan:1997kp,Lindblom:2008cm}, where SNR is the signal-to-noise ratio, this level of accuracy is already sufficient for ringdown signals with ${\rm SNR}\sim 100$--$1000$, as those expected for next-generation detectors~\cite{Bhagwat:2023jwv} and (in optimistic scenarios) for LISA~\cite{LISA:2024hlh,Berti:2016lat,Bhagwat:2021kwv}. 

Several extensions of our work are possible and desirable. The model should be validated and generalized to precessing and eccentric binaries, possibly by developing fitting formulas for the model parameters that extend the analysis of Ref.~\cite{Rosato:2025ulx}. Another important direction is the inclusion of spheroidal–spherical mode mixing~\cite{Berti:2014fga} and nonlinearities~\cite{Cheung:2022rbm,Mitman:2022qdl,Cardoso:2026llh},
which is relevant for rapidly spinning remnants and in the high-frequency regime.
It will be interesting to attach the model to inspiral–merger waveforms in the frequency domain, such as those of the IMRPhenom family~\cite{Khan:2015jqa,Pratten:2020ceb} or reduced-order frequency-domain implementations of effective-one-body models~\cite{Purrer:2015tud,Bohe:2016gbl,Gamba:2020ljo}, and to quantify the resulting overall faithfulness.
Finally, it is important to extend \textsc{GreyRing} to beyond-GR theories, complementing ordinary BH spectroscopy models beyond GR.
Some of this work is currently underway and will be presented elsewhere.

% \noindent
% {\bf \em  Software.}
Inference simulations were carried out with \texttt{pycbc inference}~\cite{Biwer:2018osg}.
Parameter estimation was performed in \texttt{Bilby}~\cite{bilby_paper}.
The manuscript content has been derived using publicly available software: {\tt matplotlib}, {\tt corner}, {\tt json}, {\tt numpy}~\cite{Hunter:2007, corner, Bray2014TheJO, harris2020array}. 
%%%%%%%%%%%%%%%%%%%%%%%%%%%%%%

%%%%%%%%%%%%%%%%%%%%%%%%%%%%%%

\noindent
{\bf \em  Acknowledgments.}
We thank Francesco Iacovelli, Xisco Jimenez Forteza, Costantino Pacilio, and Jaime Redondo-Yuste for interesting discussions. R.F.R., F.C., and P.P. are partially supported by the MUR FIS2 Advanced Grant ET-NOW (CUP:~B53C25001080001) and by the INFN TEONGRAV initiative. 
R.F.R. acknowledges the financial support provided under the ``Progetti per Avvio alla Ricerca Tipo 1'', protocol no. AR125199BD9B0F25.
S.Y. is supported by the NSF Graduate Research Fellowship Program under Grant No.~DGE2139757.
S.Y. and E.B. are supported by NSF Grants No.~AST-2307146, No.~PHY-2513337, No.~PHY-090003, and No.~PHY-20043, by NASA Grant No.~21-ATP21-0010, by John Templeton Foundation Grant No.~62840, by the Simons Foundation [MPS-SIP-00001698, E.B.], by the Simons Foundation International [SFI-MPS-BH-00012593-02], and by Italian Ministry of Foreign Affairs and International Cooperation Grant No.~PGR01167.
Some numerical computations have been performed at the Vera and CHRONOS clusters supported by the Italian Ministry of Research and by Sapienza University of Rome.
Part of this work was carried out at the Advanced Research Computing at Hopkins (ARCH) core facility (\url{https://www.arch.jhu.edu/}), which is supported by the NSF Grant No. OAC-1920103.

%%%%%%%%%%%%%%%%%%%%%%%%%%%%%%
\noindent
{\bf \em  Data Availability.}
The data supporting the findings of this study are publicly available in Ref.~\cite{GreybodyData}. 
The code used in this work is publicly available at
\href{https://github.com/romeofelicerosato-prog/GreyRing}{GreyRing}.

%%%%%%%%%%%%%%%%%%%%%
\bibliography{biblio}
%%%%%%%%%%%%%%%%%%%%%

\newpage
\appendix
\setcounter{secnumdepth}{1}
\section{graybody factor and reflection amplitude}
\label{app:gfs_theory}
Here, for completeness, we define the complex reflection amplitude
$R_{\ell m}(M\omega,\chi)$ and the associated graybody factor for gravitational perturbations of a Kerr
BH. 
In Boyer-Lindquist coordinates, the Kerr metric reads
\begin{align}
&ds^2 = -\left(1-\frac{2Mr}{\Sigma}\right)dt^2 
       + \frac{\Sigma}{\Delta}dr^2 
       - \frac{4Mar\sin^2\theta}{\Sigma}\,dt\,d\phi \notag\\
& + \Sigma\,d\theta^2
       + \left[\frac{(r^2+a^2)^2 - a^2\Delta\sin^2\theta}{\Sigma}\right]
         \sin^2\theta\,d\phi^2 ,
\end{align}
where $\Sigma=r^2+a^2\cos^2\theta$, $\Delta=r^2+a^2-2Mr$, and the horizon is
at $r_+=M+\sqrt{M^2-a^2}$.  Scalar, electromagnetic and gravitational
perturbations are described by the Teukolsky master equation~\cite{Teukolsky:1972my}.  After separation of
variables, the radial function ${}_sZ_{\ell m {\omega}}(r)$ satisfies
\begin{align}
&\Delta^{-s}\frac{d}{dr}\!\left(\Delta^{s+1}\frac{d{}_s Z_{\ell m \omega}}{dr}\right) 
+ \Bigg[\frac{K^2}{\Delta} +\quad\quad\quad\nonumber\\&\quad-\frac{2is(r-M)K}{\Delta} +4is\omega r - \lambda\Bigg]{}_s Z_{\ell m \omega}=0,
\end{align}
where $K=(r^2+a^2)\omega-am$, $s$ is the spin weight ($s=0,\pm1\pm2$ for massless scalar, electromagnetic and gravitational perturbations, respectively), and $\lambda$ is the
angular separation constant appearing in the spin-weighted spheroidal
harmonic angular equation. Introducing the tortoise coordinate $r_*$, such that $dr_*/dr=(r^2+a^2)/\Delta$,
the horizon corresponds to $r_*\to-\infty$, and spatial infinity to
$r_*\to+\infty$.  For fixed $(\ell,m,\omega)$, the physical solution is
purely ingoing at the horizon, i.e.
\begin{equation}
{}_sZ_{\ell m \omega}(r) \xrightarrow[r\to r_+]{} \Delta^{-s} e^{-ik r_*},
\qquad k=\omega-m\Omega_{\rm H} ,
\end{equation}
where $\Omega_H$ is the angular velocity of the horizon.  At large radii,
the solution behaves as
\begin{equation}
{}_sZ_{\ell m \omega}(r)\xrightarrow[r\to\infty]{}
A^{\rm out}_{\ell m {\omega}}\, \frac{e^{+i\omega r_*}}{r^{1+2s}}
+ A^{\rm in}_{\ell m {\omega}}\, \frac{e^{-i\omega r_*}}{r}\,.
\end{equation}
The amplitudes $A^{\rm in}_{\ell m \omega}$ and $A^{\rm out}_{\ell m \omega}$ are obtained by matching the
numerical solution of the Teukolsky equation with its asymptotic form through higher-order series expansions~\cite{Pani:2013pma}.  
Since the amplitude of the Teukolsky variable is not directly associated with the energy flux, one must convert
$(A^{\rm in}_{\ell m \omega},A^{\rm out}_{\ell m \omega})$ into physical reflection and transmission
amplitudes.  The required normalization factors were derived in~\cite{Press:1972zz,Press:1973zz,Teukolsky:1974yv} and independently in~\cite{Starobinskii:1973vzb,Starobinskil:1974nkd}.  For spin $s=-2$
gravitational perturbations, the reflection amplitude is given by 
\begin{equation}
R_{\ell m}( M\omega,\chi)
  = \frac{C}{\,(2M\omega)^4}
 \frac{A^{\rm out}_{\ell m\omega}}{A^{\rm in}_{\ell m\omega}} ,
\end{equation}
where 
\begin{align}
|C|^2 &= B\Big[\big(\lambda + s(s+1) -2\big)^{2}
              + 36 a\omega m
              - 36 a^{2}\omega^{2}\Big]
\nonumber\\
&\quad + \left[2\big(\lambda + s(s+1)\big) -1\right]
       \left(96 a^{2}\omega^{2} - 48 a\omega m\right)
\nonumber\\
&\quad + 144\,\omega^{2}(M^{2}-a^{2})\,,\\
{\rm Im}[C]&=12 M\omega\nonumber\\
\nonumber
B &= \left(\lambda + s(s+1)\right)^{2}
     + 4 m a \omega
     - 4 a^{2}\omega^{2}\,,\nonumber
\end{align}
and the real part of $C$ is positive~\cite{Teukolsky:1974yv}.
These coefficients encode the flux normalization appropriate for the
Teukolsky radial function, ensuring that the reflection probability $\mathcal{R}_{\ell m}( M\omega,\chi)=|R_{\ell m}( M\omega,\chi)|^2=1$
corresponds to perfect reflection, while $\mathcal{R}_{\ell m}( M\omega,\chi)=0$ corresponds to perfect absorption by the BH.

Finally, for each $(\ell,m)$, the graybody factor $\Gamma_{\ell m}(M\omega,\chi)$ is connected to the reflection  probability through 
\begin{equation}\label{eq:gfsrefl}
\mathcal{R}_{\ell m}(M\omega,\chi)=1-{\omega-m\Omega_{\rm H} \over\omega}\Gamma_{\ell m}(M\omega,\chi)\,,
\end{equation}
which encodes the superradiant amplification when $0<\omega<m\Omega_{\rm H}$~\cite{Brito:2015oca}.

%%%%%%%%%%%%%%%%%%%%%%%%%%%%%%%%%%
\section{graybody-factor waveform model and fits to numerical simulations} \label{app:fittingprocedure}
%%%%%%%%%%%%%%%%%%%%%%%%%%%%%%%%%%
Here we provide further details on the \textsc{GreyRing} model and on the fitting procedure to numerical relativity simulations.

\subsection{\textsc{GreyRing} model}

Recently, Ref.~\cite{Rosato:2026moe} demonstrated that the reflection amplitude $R_{\ell m}(M\omega,\chi)$ is explicitly encoded in the component of the Green's function governing the linear ringdown response of a BH in the frequency domain. This result provides analytical support to earlier phenomenological models of the spectral amplitude of the signal~\cite{Oshita:2023cjz,Rosato:2024arw,Rosato:2025byu,Rosato:2025ulx}.
In particular, Ref.~\cite{Rosato:2025ulx} showed that a model for the spectral amplitude of the form
\begin{equation}\label{eq:ampmodel}
|\tilde h_{\ell m} (\omega)|= \frac{M A_{\ell m}}{(M\omega)^{p_{\ell m}}}\,|R_{\ell m}(M\omega, \chi)|
\end{equation}
provides an accurate fit to numerical waveforms, but it did not include a model for the phase of the signal.

Motivated by this parametrization, and by the fact that the full signal must be proportional to the complex reflection amplitude $R_{\ell m}(M\omega,\chi)$~\cite{Rosato:2026moe}, we consider the following theory-informed model for the full frequency-domain signal:
\begin{equation}\label{eq:fullmodelB}
\tilde h_{\ell m} (\omega)= \frac{M A_{\ell m}}{(M\omega)^{p_{\ell m}}}\,R_{\ell m}(M\omega, \chi)\,e^{i \frac{c_{\ell m}}{(M\omega)^{d_{\ell m}}}} \, e^{i(\phi_c + \omega t_c)}\,.
\end{equation}
The phase term $c_{\ell m}/(M\omega)^{d_{\ell m}}$ plays a role analogous to the amplitude correction $A_{\ell m}/(M\omega)^{p_{\ell m}}$.

We have explored different modeling choices for the phase. We first considered a minimal model in which the phase is entirely determined by $\arg(R_{\ell m})$ (i.e., we set $c_{\ell m}=0$ in Eq.~\eqref{eq:fullmodelB}), supplemented by the extrinsic parameters $\phi_c$ and $t_c$. Remarkably, this simple prescription already provides a good description of the numerical phase, with typical mismatches of order $\mathcal{O}(10^{-3})$ for the $\ell=m=2$ mode. This indicates that the phase of the reflection amplitude captures the dominant physical behavior of the signal.
We then extended the model by including subleading corrections of the form $\sim c_{\ell m}/(M\omega)^{d_{\ell m}}$ and, finally, the simplified expression $\sim c_{\ell m}/(M\omega)$. Both extensions lead to a significant improvement in the mismatch, reducing it by up to three orders of magnitude. 

\begin{figure}[th]
    \centering
    \includegraphics[width=\linewidth]{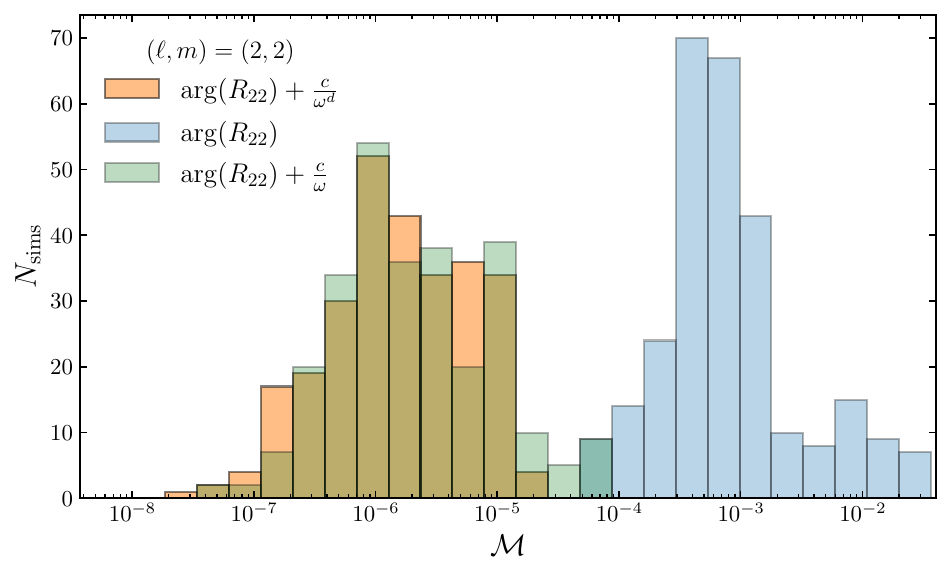}
    \caption{Distribution of the mismatch $\mathcal{M}$ for the $(\ell,m)=(2,2)$ mode of \textsc{GreyRing} over $311$ comparable-mass, spin-aligned binary BH simulations from the SXS catalog~\cite{Boyle:2019kee,Scheel:2025jct}. The baseline model $\arg(R_{22})$ (blue) already provides a good description of the data, with mismatches clustered around $\sim 10^{-4}$--$10^{-3}$. Including a correction of the form $c_{\ell m}/\omega^{d_{\ell m}}$ (orange) significantly improves the agreement, shifting the distribution to lower values. Fixing $d_{\ell m}=1$ (green) yields nearly identical results, indicating that the simplified $c_{\ell m}/(M\omega)$ parametrization captures the relevant subleading phase structure without loss of accuracy.}
    \label{fig:mismatchdist}
\end{figure}

These models are compared in Fig.~\ref{fig:mismatchdist} 
against $311$ comparable-mass, quasicircular, nonprecessing binary BH simulations from the Simulating eXtreme Spacetimes~(SXS) collaboration catalog~\cite{Boyle:2019kee,Scheel:2025jct}. The baseline model already achieves relatively small mismatches, confirming that $\arg(R_{\ell m})$ captures most of the phase structure. The inclusion of the $c_{\ell m}/(M\omega)^{d_{\ell m}}$ correction leads to a significant improvement, reducing the mismatch distribution by about three orders of magnitude. Allowing the exponent $d_{\ell m}$ to vary as a free parameter does not yield significant improvement in the mismatch: the best-fit values are consistently clustered around $d_{\ell m}\simeq 1$, and fixing $d_{\ell m}=1$ yields nearly indistinguishable mismatches. For this reason, in the main text and below we fix $d_{\ell m}=1$ and we adopt the simpler $c_{\ell m}/(M\omega)$ parametrization, which retains the accuracy of the more general parametrization while reducing the number of free parameters. The $c_{\ell m}/(M\omega)$ correction can be understood as the leading-order high-frequency contribution in a Laurent expansion of the phase.

%%%%%%%%%%%%%%%%%%%%%%%%%%%%%%%%
\subsection{Fits to numerical simulations}
%%%%%%%%%%%%%%%%%%%%%%%%%%%%%%%%
Here we describe the fitting procedure used to compare and validate \textsc{GreyRing} against numerical relativity simulations. 
The latter encompass binaries with mass ratio $q\in [1,4]$. We neglect the oldest (and presumably less accurate) simulations of the SXS catalog, focusing on simulation ID$\geq1000$. As in Ref.~\cite{Rosato:2025ulx}, we Fourier-transform these simulations using the preprocessing routines implemented in the \texttt{sxs} Python package; the need for this preprocessing is discussed in Ref.~\cite{Chen:2024ieh}.

The frequency-domain waveform is modeled as 
\begin{equation}\label{eq:SMfullmodel}
\tilde h_{\ell m} (\omega)= {MA_{\ell m} \over (M\omega)^{p_{\ell m}}}\,R_{\ell m}(M\omega, \chi)\,e^{i {c_{\ell m} \over M\omega}} e^{i(\phi_c + \omega t_c)}\,,
\end{equation}
and the multipolar components $\tilde{h}_{\ell m}$ are computed using the Fourier-transform convention used in the main text and in Eq.~\eqref{eq:FT_main} below. 

For given values of $M$ and $\chi$, the amplitude parameters $A_{\ell m}$ and $p_{\ell m}$ are determined following the procedure introduced in Ref.~\cite{Rosato:2025ulx}, here extended to include the additional phase parameter $c_{\ell m}$.
We perform separate fits for the amplitude and phase, using the absolute value and argument of the waveform defined in Eq.~\eqref{eq:SMfullmodel}. As discussed in the main text, the phase contains two arbitrary constants, $\phi_c$ and $t_c$. To remove their effect, we detrend both the numerical phase and the model by subtracting a best-fit linear function in frequency, ensuring that the fit is not dominated by these unphysical degrees of freedom. Residual linear contributions may persist after detrending; we therefore perform the final fit using the full expression in Eq.~\eqref{eq:SMfullmodel}.
The fit is performed over a frequency interval $[\omega_i,\omega_f]$ defined as follows~\cite{Rosato:2025ulx}:
\begin{itemize}
    \item We set $\omega_i = 0.7\,\omega_x$, where $\omega_x$ is the knee frequency at which the Fourier-domain amplitude transitions from a power-law behavior to an exponential decay, approximately marking the onset of ringdown.
    \item We define $\omega_f$ such that $|\tilde{h}(\omega_f)| = |\tilde{h}(\omega_x)|/20$, in order to suppress contamination from numerical noise at high frequencies, as discussed in Ref.~\cite{Rosato:2025ulx}.
\end{itemize}
The lower cutoff $\omega_i$ is chosen so as to ensure a stable extraction of the model parameters. Further details are provided in the following subsection and in Ref.~\cite{Rosato:2025ulx}. While this condition is, strictly speaking, simulation-dependent, we find empirically that choosing $\omega_i = 0.7\,\omega_x$ consistently satisfies this requirement across the dataset.
The upper cutoff $\omega_f$ is set such that the high-frequency numerical-noise plateau of the Fourier-transformed SXS waveform is not larger than $\sim 10^{-4}$ of the signal amplitude, preventing contamination of the fit and avoiding spurious constraints on the parameters.
We have checked that the fit and mismatch are robust against variations of the frequency range.

\begin{figure}[t]
    %\centering
    \includegraphics[width=0.48\textwidth]{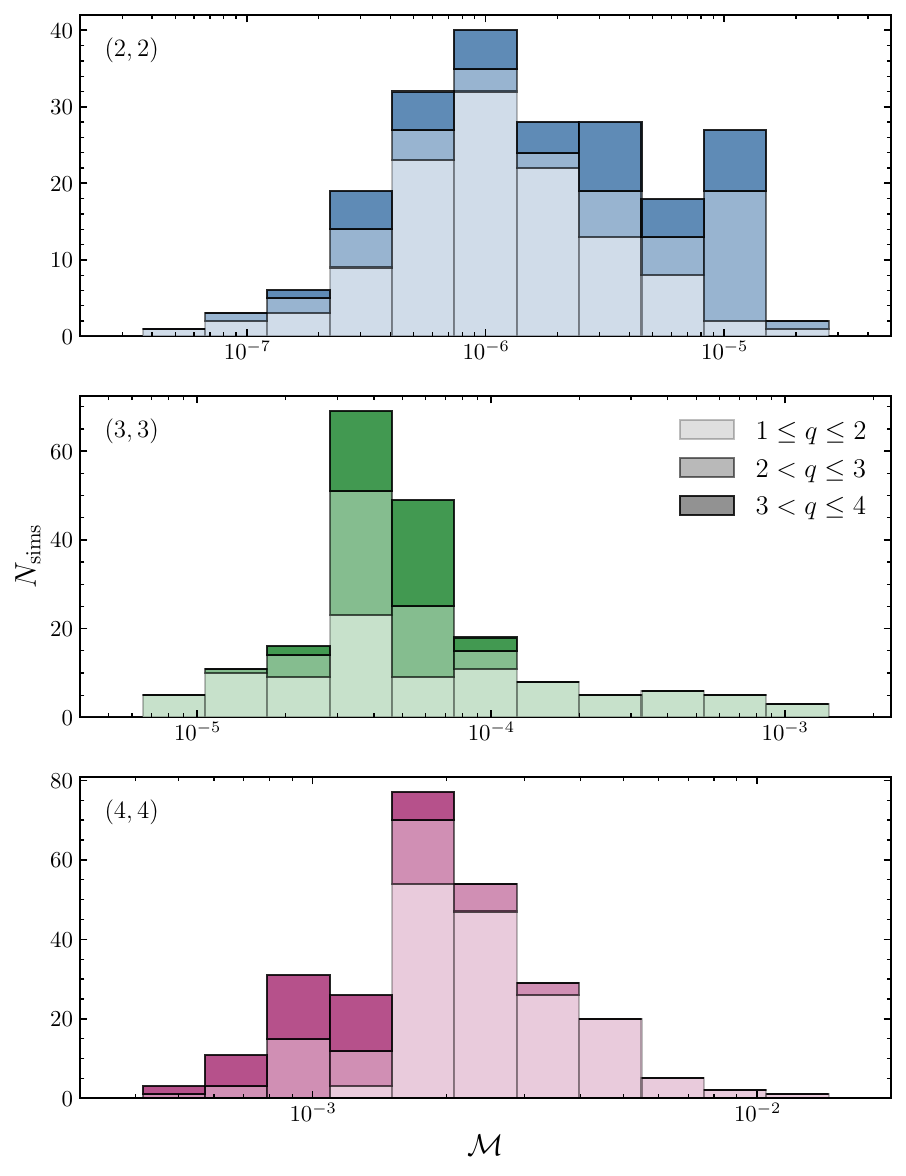}
    \caption{
Distributions of the mismatch $\mathcal{M}$ for the $(\ell,\,m)=(2,\,2),\,(3,\,3)$ and $(4,\,4)$ multipoles, split according to the mass ratio $q$.
For the dominant $(2,\,2)$ mode, equal-mass systems ($q\simeq 1$) yield the best agreement, with a sharp peak around $\mathcal{M}\sim10^{-6}$, while higher mass ratios are clustered at $\mathcal{M}\sim10^{-5}$.
For the subdominant $(3,\,3)$ and $(4,\,4)$ multipoles, the trend is reversed: higher mass ratios ($q>2$) lead to systematically smaller mismatches, whereas comparable-mass systems ($1\le q\le 2$) exhibit larger values of $\mathcal{M}$.
This behavior reflects the suppression of odd-$m$ multipoles in the comparable-mass limit, that makes them more susceptible to numerical noise and mode mixing, while they become better resolved for unequal-mass binaries. 
}
\label{fig:fit_higher_modes}
\end{figure}

We first apply the model to the dominant $(\ell,m)=(2,2)$ mode, finding excellent agreement with the numerical data, as shown in the main text and in the top panel of Fig.~\ref{fig:fit_higher_modes}. The mismatches are typically in the range $\sim 10^{-7}$--$10^{-5}$, with a distribution peaked around $\sim 10^{-6}$, demonstrating that the graybody-based description captures both the amplitude and phase of the ringdown signal with remarkable accuracy. 
We then extend the analysis to subdominant multipoles, focusing on $(\ell,m)=(3,3)$ and $(4,4)$. Despite their lower intrinsic amplitudes
and increased sensitivity to numerical artifacts, the model continues to provide an accurate description, as shown in the middle panel of Fig.~\ref{fig:fit_higher_modes}. 

The mismatch for the $(3,3)$ mode is typically in the range $10^{-5}$--$10^{-4}$, whereas for the $(4,4)$ mode (bottom panel of Fig.~\ref{fig:fit_higher_modes}) it is of order $10^{-3}$.
This is expected because of the stronger impact of numerical noise, the lower amplitude of higher multipoles, and possible spherical-spheroidal mode mixing at higher frequencies~\cite{Berti:2014fga}, as well as potential nonlinear effects~\cite{Cheung:2022rbm,Mitman:2022qdl,Berti:2025hly}. Since the $(3,3)$ mode vanishes in the equal-mass limit~\cite{Buonanno:2006ui,Berti:2007fi}, we restrict the analysis to mass ratios $q>1.001$ for this multipole.
When splitting the results by mass ratio, a clear trend emerges across all modes. For the dominant $(2,\,2)$ mode, equal-mass systems ($q\simeq 1$) yield the smallest mismatches, with a sharp peak around $\mathcal{M}\sim10^{-6}$, while higher mass ratios are clustered around $\sim10^{-5}$. For the subdominant multipoles with $(\ell,\,m)=(3,\,3)$ and $(4,\,4)$, the trend is reversed: larger asymmetries ($q>2$) lead to systematically smaller mismatches, whereas comparable-mass systems ($1\le q\le 2$) exhibit larger values of $\mathcal{M}$. This behavior reflects the suppression of odd-$m$ multipoles in the comparable-mass limit, where they become more susceptible to numerical noise and mode mixing, while they are better resolved for unequal-mass binaries.

Note that the reflectivity is naturally defined in a spheroidal-harmonic basis, whereas gravitational-wave signals and numerical relativity waveforms are typically decomposed in spherical harmonics. Transforming between these two bases induces mode mixing, with coefficients that depend on $a\omega$~\cite{Berti:2005gp,Berti:2014fga}. In this analysis we neglect such mixing, as it is expected to be subdominant in the frequency range of interest ($a\omega\lesssim 0.5$), but its impact may be nonnegligible for other modes. We defer a more careful treatment of mode mixing to future work.
%%%%%%%%%%%%%%%%%%%%%%%%%%%%%%%%%%
\subsection{Choice of the initial frequency}\label{sec:omegai}
To determine the lower frequency cutoff $\omega_i$ at which the model becomes valid, we follow the strategy of Ref.~\cite{Rosato:2025ulx}. The idea is that, within the regime of validity of the model, small changes in the initial fitting frequency should not significantly affect the inferred parameters. Thus, the parameters $(A_{\ell m},\,p_{\ell m},\,c_{\ell m})$ should display a stable plateau as $\omega_i$ is varied.

We first consider the dominant multipole, $(\ell,|m|)=(2,2)$. As a reference scale, we use the frequency $\omega_x$ introduced in the previous subsection. Since $\omega_x$ is numerically close to the fundamental QNM frequency of the remnant, using alternatively the latter as a reference scale would lead to equivalent results for the present purpose. We then scan over different values of $\omega_i$ and identify the range in which the extracted values of $(A_{\ell m},\,p_{\ell m},\,c_{\ell m})$ remain stable. Specifically, we define the stability region as the range of initial frequencies for which all three parameters vary by less than $0.1\%$.

\begin{figure}[t]
\centering
\includegraphics[width=\columnwidth]{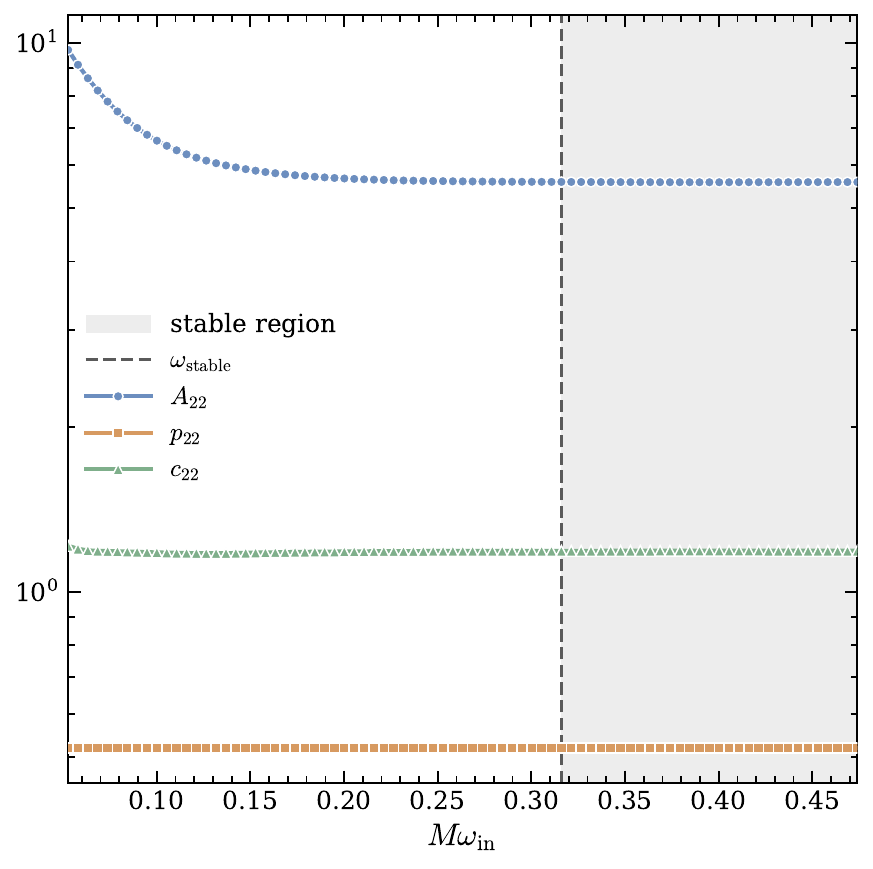}
\caption{Stability of the extracted model parameters as a function of the lower fitting frequency $M\omega_i$ for the dominant multipole $(\ell,m)=(2,2)$ of SXS:BBH:3977. The parameters $(A_{22},\,p_{22},\,c_{22})$ are obtained by repeating the fit while varying $\omega_i$. The gray shaded region marks the range in which all three parameters vary by less than $0.1\%$, which we identify as the stability region. For this simulation the onset of stability occurs at $M\omega_i\simeq0.32$, corresponding to $\omega_i\simeq0.6\omega_x$.}
\label{fig:stability_3977}
\end{figure}

An example of this procedure is shown in Fig.~\ref{fig:stability_3977} for the simulation \textsc{SXS:BBH:3977}. The gray shaded region denotes the stability region according to the criterion defined above. In this case, stability is reached at approximately $M\omega_i \simeq 0.32$, corresponding to $\omega_i \simeq 0.6\,\omega_x$. We note that the extracted parameters appear stable already at slightly lower frequencies. Nevertheless, we adopt the quantitative criterion above to conservatively define the onset of the stable regime.

\begin{figure}[t]
\centering
\includegraphics[width=\columnwidth]{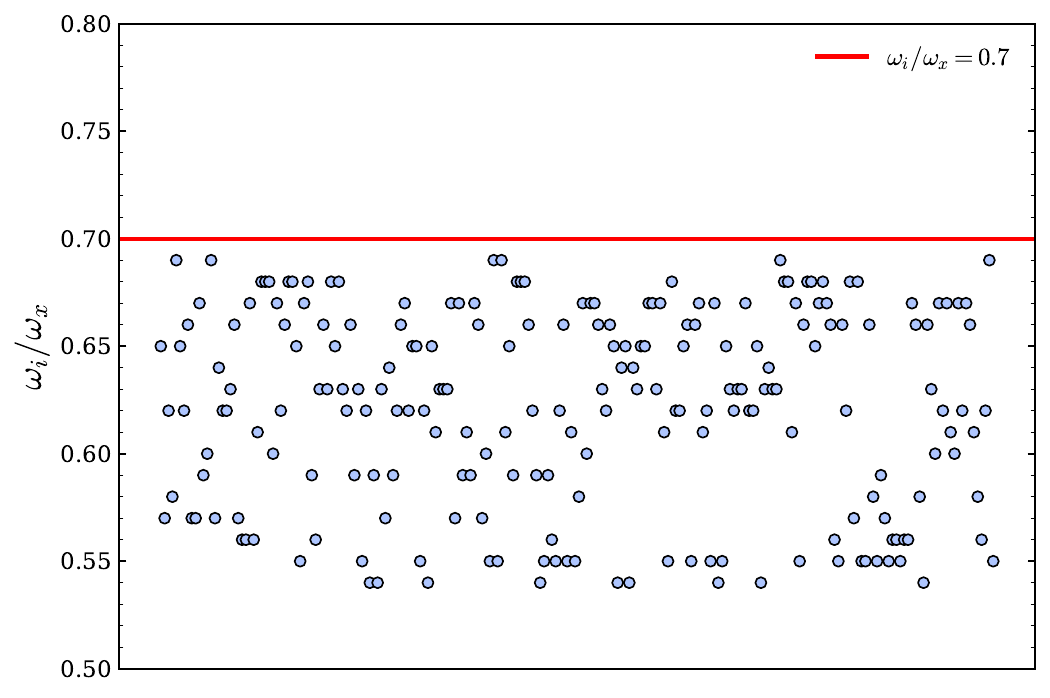}
\caption{
Stable lower cutoff frequencies selected by the stability criterion for the numerical relativity simulations considered in this work. Each point corresponds to one simulation and shows the cutoff in units of the reference frequency $\omega_x$; the ordering along the horizontal axis is only for visualization. For the dominant multipole $(\ell,|m|)=(2,2)$, the onset of stability lies in the range $\omega_i/\omega_x\simeq0.55$--$0.70$ for all simulations. The horizontal red line marks the conservative choice $\omega_i=0.7\omega_x$ adopted in the rest of the analysis.
}
\label{fig:omega_i_histogram}
\end{figure}

We apply the same procedure to a subset of the numerical relativity simulations analyzed in the previous section, restricting to binaries with individual spin magnitudes below $0.8$. The resulting stable lower cutoffs are shown in Fig.~\ref{fig:omega_i_histogram}: each point corresponds to one simulation and gives the value of $\omega_i/\omega_x$ at which the stability criterion is first satisfied. The ordering along the horizontal axis is only for visualization. The onset of the stable regime typically falls in the range
$0.55 \lesssim \omega_i/\omega_x \lesssim 0.70$.
Thus, the horizontal red line at $\omega_i/\omega_x=0.7$ provides a conservative upper bound for the simulations considered here. However, individual simulations may allow different choices of $\omega_i$, and a dedicated stability analysis can be performed case by case when needed.

Finally, we apply the same stability procedure to the higher multipoles. The method is identical to that described above and in Ref.~\cite{Rosato:2025ulx}, except that the stability threshold is mildly relaxed for modes with smaller Fourier-domain amplitudes. Since we mainly consider comparable-mass mergers, the higher modes are typically less pronounced than the dominant $(2,2)$ mode. We therefore use a stability threshold of $1\%$ for the $(3,3)$ and $(4,4)$ modes. This preserves the same operational definition of a stable extraction while accounting for the reduced numerical accuracy of subdominant modes.

The same underlying idea can be extended to real gravitational-wave data, although its practical implementation requires a modified procedure, discussed below.

%%%%%%%%%%%%%%%%%%%%%%%%%%%%%%%%%%
\subsection{Remnant mass and spin determination}

A central requirement for using \textsc{GreyRing} as a consistency test of GR is that the model recover remnant parameters $(M,\chi)$ consistent with numerical relativity, rather than merely reproduce the waveform morphology. As a first direct check, we perform two-dimensional scans over the remnant mass and spin that appear in the graybody factor $R_{22}(M\omega,\chi)$, and refitting the remaining model parameters at each grid point. 
We consider three representative nonprecessing SXS simulations, SXS:BBH:1137, SXS:BBH:3617, and SXS:BBH:3917, with remnant spins $\simeq(0.369, 0.686, 0.877)$, probing low-, intermediate-, and high-spin remnants, respectively. For each simulation, we compute the \textsc{GreyRing} mismatch in the $(M,\chi)$ plane, minimizing over the phenomenological parameters $(A_{\ell m},\,p_{\ell m},\,c_{\ell m})$ and the extrinsic time and phase shifts $(t_c,\phi_c)$.

\begin{figure}
\centering
\includegraphics[width=0.5\textwidth]{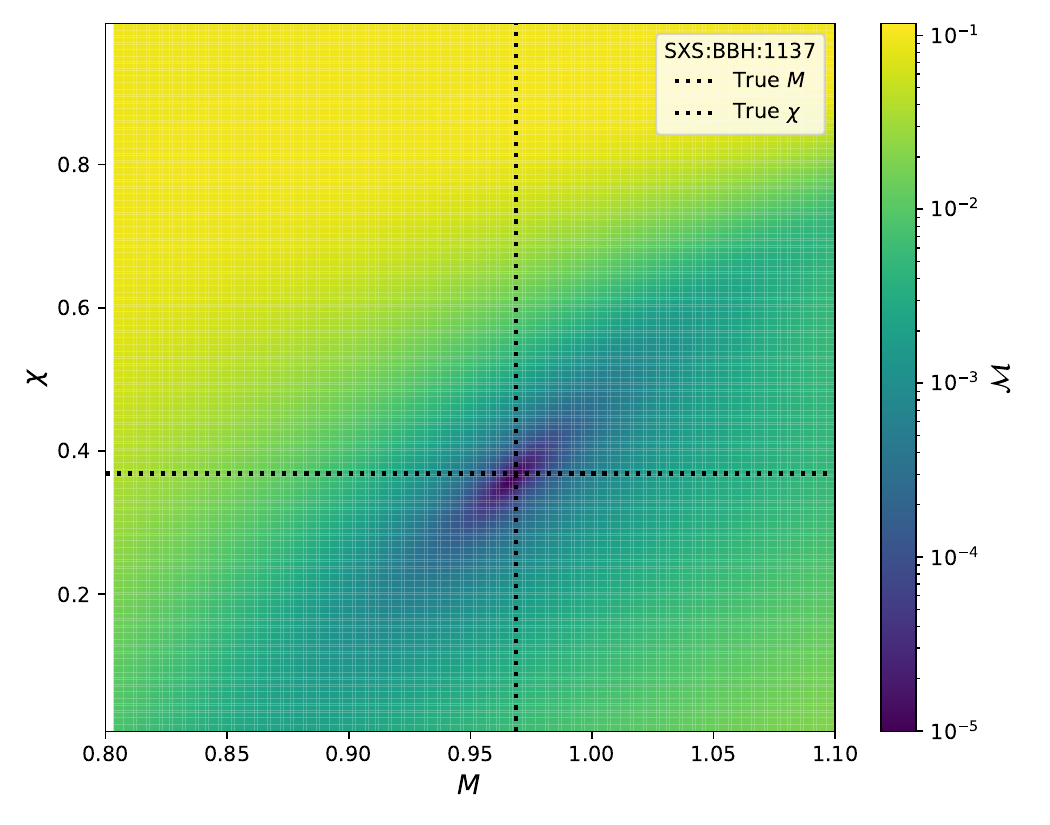}
\includegraphics[width=0.5\textwidth]{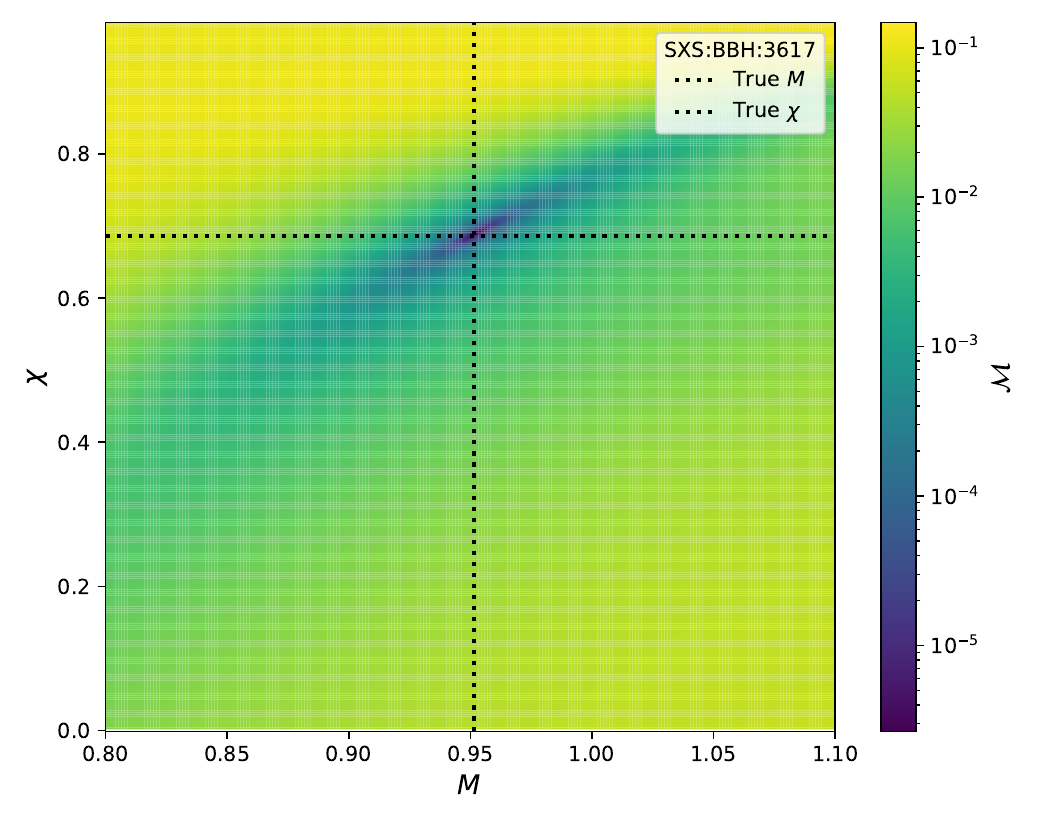}
\includegraphics[width=0.5\textwidth]{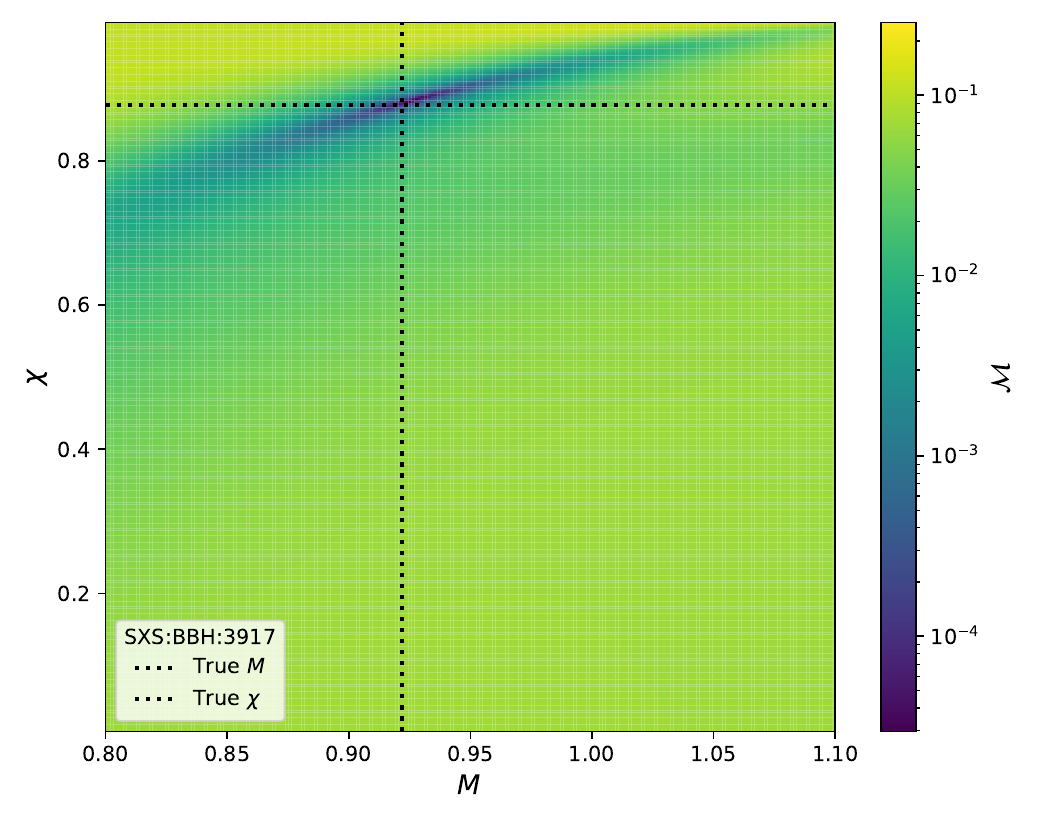}
\caption{
Remnant mass and spin recovery for three representative SXS simulations. We show the \textsc{GreyRing} mismatch in the $(M,\chi)$ plane for SXS:BBH:1137, SXS:BBH:3617, and SXS:BBH:3917, whose numerical relativity remnant spins are $\chi\simeq(0.369,0.686,0.877)$, respectively. At each grid point, the remaining \textsc{GreyRing} parameters are refitted. The black dotted lines indicate the numerical relativity remnant values. In all three cases, the numerical relativity prediction lies within the minimum-mismatch region, demonstrating that the \textsc{GreyRing} fit correctly recovers the remnant Kerr geometry.}\label{fig:mass_spin_recovery}
\end{figure}

The resulting maps are shown in Fig.~\ref{fig:mass_spin_recovery}. In all three cases, the numerical relativity remnant values lie within the minimum-mismatch region, with the low-mismatch basin centered around the numerical relativity prediction despite the different final spins. This indicates that the preferred values of $M$ and $\chi$ are selected by the mass- and spin-dependent structure of the graybody factor. These scans therefore provide a direct validation of the interpretation of $M$ and $\chi$ as physical remnant parameters in the \textsc{GreyRing} model.

Motivated by these results, we apply the same fitting strategy to a larger set of numerical relativity simulations, in order to test whether the remnant mass and spin are systematically recovered by the model. To quantify the agreement between the numerical relativity remnant parameters and the values reconstructed by the \textsc{GreyRing} fit, we follow the approach of Ref.~\cite{Giesler:2019uxc} and define
\begin{equation}
\epsilon =
\sqrt{
\left(\frac{M_{\rm fit}-M_{\rm NR}}{M_{\rm NR}}\right)^2
+
\left(\chi_{\rm fit}-\chi_{\rm NR}\right)^2
} .
\label{eq:epsilon_remnant}
\end{equation}
Before performing this comparison, we detrend the phase by removing the best-fit extrinsic time and phase shifts $(t_c,\phi_c)$, which carry no information about the remnant geometry. We then scan a grid in remnant mass and spin, refitting the model and computing the mismatch at each grid point. Finally, we refine the scan around the minimum-mismatch region to determine the best-fit values $(M_{\rm fit},\chi_{\rm fit})$.

\begin{figure}
\centering
\includegraphics[width=0.45\textwidth]{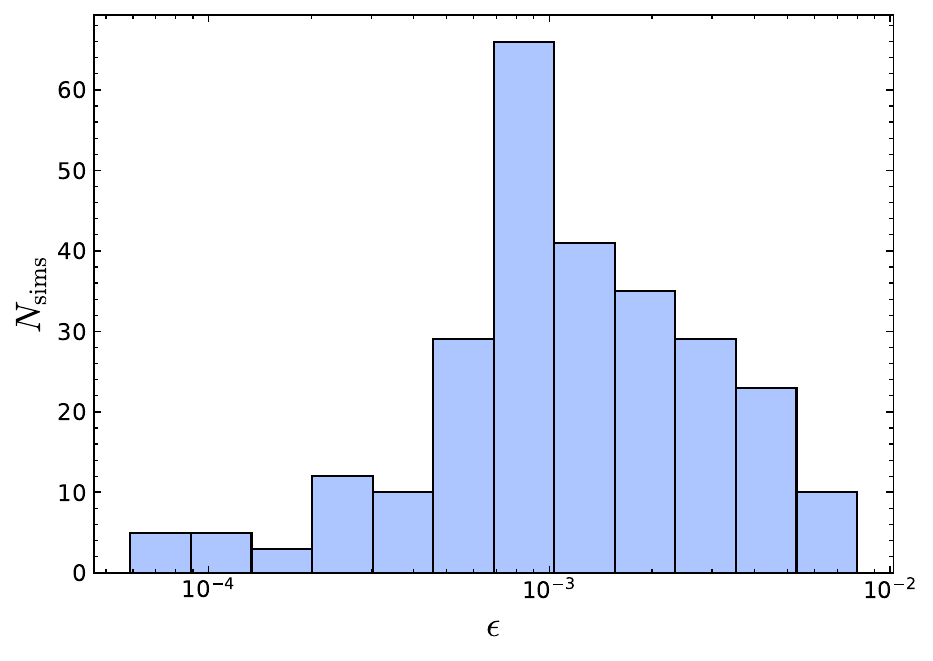}
\caption{
Distribution of the remnant-parameter reconstruction error $\epsilon$, defined in Eq.~\eqref{eq:epsilon_remnant}, for the subset of quasicircular, nonprecessing SXS simulations with $\mathrm{ID}>1000$ and $q<4$. The distribution peaks at $\epsilon\sim10^{-3}$ and remains below $10^{-2}$ for all simulations considered here, with several cases reaching values of order $10^{-4}$ or smaller. This level of accuracy is consistent with the remnant-parameter reconstruction obtained from QNM fits in Ref.~\cite{Giesler:2019uxc}.}
\label{fig:epsilon_remnant_hist}
\end{figure}

The resulting values of $\epsilon$ are shown in Fig.~\ref{fig:epsilon_remnant_hist}. The distribution peaks at $\epsilon\sim10^{-3}$, and all simulations considered here satisfy $\epsilon<10^{-2}$, with several reaching values of order $10^{-4}$ or smaller. This level of accuracy is fully consistent with the remnant-parameter reconstruction obtained in Ref.~\cite{Giesler:2019uxc} using QNMs. 
The peak at $\epsilon\sim10^{-3}$ may also partly reflect the numerical resolution of the graybody-factor tables employed in the fit, which are sampled on a grid in $(M\omega,\chi)$ with spacing of order $10^{-3}$.

%%%%%%%%%%%%%%%%%%%%%%%%%%%%%%%%%%
\subsection{Extrinsic parameters and angular dependence}\label{app:angles}

In the main text we adopt the Fourier convention
\begin{equation}
\tilde h(\omega)=\int_{-\infty}^{+\infty} h(t)\,e^{i\omega t}\,dt,
\label{eq:FT_main}
\end{equation}
under which, for nonprecessing binaries, modes with $m>0$ are concentrated at positive frequencies, while modes with $m<0$  are concentrated at negative frequencies~\cite{Garcia-Quiros:2020qpx}. 

For the implementation in gravitational-wave data analysis, we instead use the LIGO Algorithms Library~(LAL)~\cite{lalsuite} convention,
\begin{equation} \label{eq:FT_LAL}
\tilde h(f)=\int_{-\infty}^{+\infty} h(t)\,e^{-2\pi i f t}\,dt,
\end{equation}
which is also used in \texttt{Bilby}~\cite{bilby_paper}. Equations~\eqref{eq:FT_LAL} and~\eqref{eq:FT_main} are related by $\tilde h(f)=\tilde h(\omega=-2\pi f)$. In the second convention the roles of the two branches are reversed: modes with $m>0$ are concentrated predominantly at negative frequencies, while modes with $m<0$ are concentrated at positive frequencies. 
Therefore, for $f>0$, the relevant contribution is provided by $\tilde h_{\ell\,-m}(f)$.

The equatorial symmetry of nonprecessing binaries implies
\begin{equation}
\tilde h_{\ell m}(f)=(-1)^\ell \tilde h^*_{\ell\,-m}(-f),
\end{equation}
so that the full signal is determined by the positive-frequency modes.

The detector-frame polarizations in the time domain are obtained from the multipolar decomposition
\begin{equation}
h_+ - i h_\times = \frac{1}{d_L}\sum_{\ell m} h_{\ell m}\,{}_{-2}Y_{\ell m}(\theta,\phi),
\end{equation}
which, for positive frequencies and using the symmetry above, yields
\begin{align}
\tilde h_{+,\,l|m|}(f) &= \frac{1}{2d_L}\left({}_{-2}Y_{\ell\,-m} + (-1)^\ell {}_{-2}Y^*_{\ell m}\right)\tilde h_{\ell\,-m}(f), \\
\tilde h_{\times,\,l|m|}(f) &= \frac{i}{2d_L}\left({}_{-2}Y_{\ell\,-m} - (-1)^\ell {}_{-2}Y^*_{\ell m}\right)\tilde h_{\ell\,-m}(f),
\end{align}
valid for $f>0$.

Restricting to the dominant quadrupolar contribution $(\ell,m)=(2,\pm2)$, and using $(-1)^\ell=1$, the relevant spin-weighted spherical harmonics are
\begin{align}
{}_{-2}Y_{2\,\pm2}(\theta,\phi) &= \sqrt{\frac{5}{64\pi}}(1\pm\cos\theta)^2 e^{\pm2i\phi}\,.
\end{align}

Substituting into the expressions above, one obtains
\begin{align}
\tilde h_+(f) &= \sqrt{\frac{5}{4\pi}}\,\frac{1+\cos^2\theta}{2}\,\frac{e^{-2i\phi}}{2d_L}\,\tilde h_{2\,-2}(f), \\
\tilde h_\times(f) &= -\,i\,\sqrt{\frac{5}{4\pi}}\,\cos\theta\,\frac{e^{-2i\phi}}{2d_L}\,\tilde h_{2\,-2}(f).
\end{align}

We therefore define the effective positive-frequency mode
\begin{equation}
\frac{e^{-2i\phi}}{2d_L}\,\tilde h_{2\,-2}(f)
\equiv
\frac{H(f)}{2}\,e^{-i\Phi(f)},
\end{equation}
so that the polarizations reduce to the standard form
\begin{align}
\tilde h_{+,\,2|2|}(f) &= \sqrt{\frac{5}{4\pi}}\,\frac{1+\cos^2\theta}{2}\,\frac{H(f)}{2}\,e^{-i\Phi(f)}, \\
\tilde h_{\times,\,2|2|}(f) &= -\,i\,\sqrt{\frac{5}{4\pi}}\,\cos\theta\,\frac{H(f)}{2}\,e^{-i\Phi(f)}.
\end{align}
The term $\Phi(f)$ corresponds to the phase defined in Eq.~\eqref{eq:SMfullmodel}, with an overall minus sign arising from the different Fourier transform convention adopted here. The quantity $H(f)$ is related to Eq.~\eqref{eq:ampmodel} once physical units are restored, namely 
\begin{equation}
H(f)= A_{22}\left(\frac{t_M^2 c^2}{d_L}\right)\, \,
\frac{R_{22}\!\left(2\pi t_M f,\chi\right)}
{\left(2\pi t_M f\right)^{p_{22}}}\,,
\end{equation}
where $t_M=G M/c^3$.

%%%%%%%%%%%%%%%%%%%%%%%%%%%%%%%%%%
\subsection{Mismatch on the sphere}
\begin{figure}[t]
    \centering
    \includegraphics[width=0.5\textwidth]
    {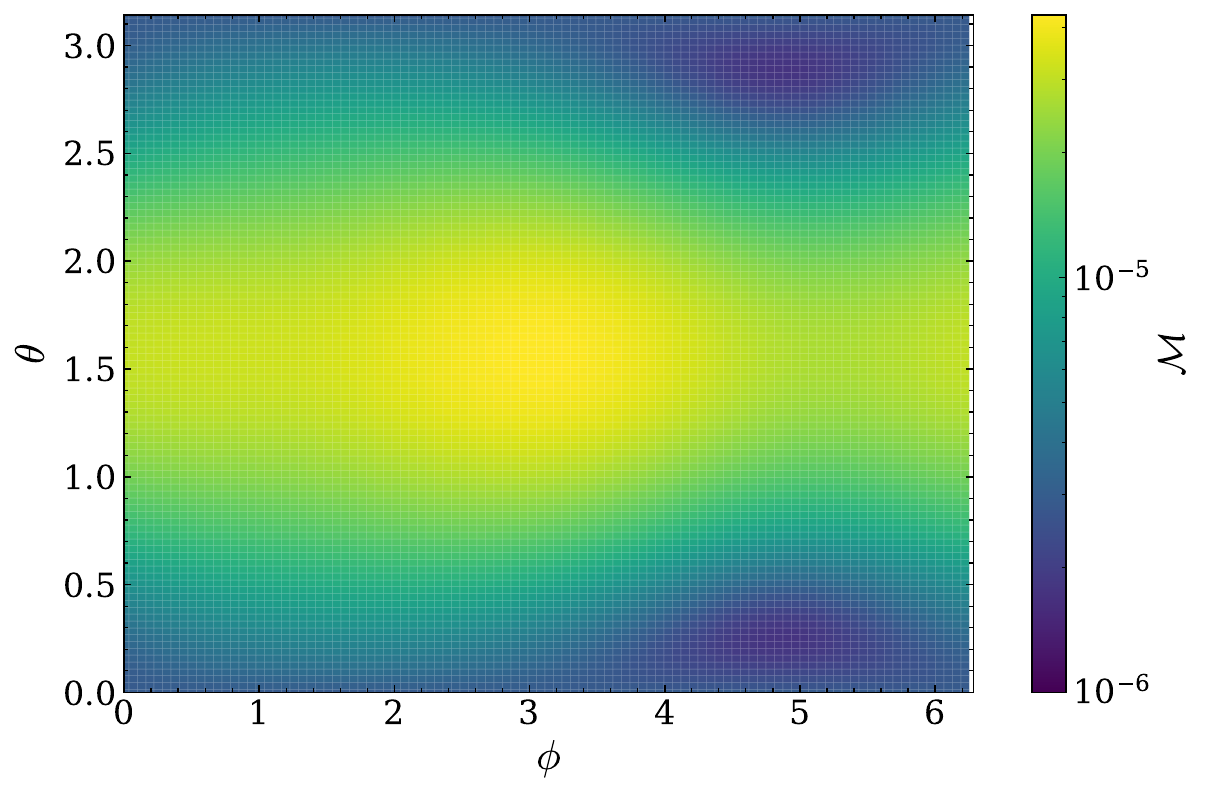}
    \caption{
        Mismatch over the sphere for \textsc{SXS:BBH:3977}, obtained by comparing the summed \textsc{GreyRing} waveform, including the $(2,\pm2)$, $(3,\pm3)$, and $(4,\pm4)$ modes, with the corresponding numerical relativity waveform truncated to the same set of modes. The mismatch is shown as a function of the orientation angles $(\theta,\phi)$. It is typically of order $10^{-6}$ over most of the sphere and increases to values of order $10^{-5}$ near $\theta\sim\pi/2$. Despite the larger individual mismatches of the subdominant $(3,3)$ and $(4,4)$ modes, the total mismatch remains close to that of the dominant $(2,2)$ contribution.
    }
    \label{fig:mismatch_sphere_3977}
\end{figure}

%%%%%%%%%%%%%%%%%%%%%%%%%%%%%%%%%%
So far, we have discussed the mismatches of the individual multipoles. We now assess how these mode-by-mode discrepancies affect the mismatch of the full observable signal across the sphere. We restrict the analysis to the modes modeled so far, namely $(\ell,m)=(2,\pm2)$, $(3,\pm3)$, and $(4,\pm4)$. We therefore compare the summed \textsc{GreyRing} waveform with the numerical relativity waveform truncated to the same set of modes.
This provides a more direct measure of how the individual-mode mismatches affect the total signal observed at a given orientation $(\theta,\phi)$. As an illustrative example, Fig.~\ref{fig:mismatch_sphere_3977} shows the mismatch across the sphere for the simulation \textsc{SXS:BBH:3977}, represented in the $(\theta,\phi)$ plane.
At each orientation, we compute the unweighted mismatch of the complex strain over the same frequency interval used for the individual-mode analysis, without further optimization over orientation-dependent time or phase shifts.

The mismatch ranges from $10^{-6}$ to a few times $10^{-5}$ across the sphere. In particular, it is of order $10^{-6}$ for nearly face-on and face-off orientations, while reaching values of order $10^{-5}$ for edge-on systems. This behavior is consistent with the fact that, although the individual $(3,3)$ and $(4,4)$ modes have mismatches of order $10^{-4}$ and $10^{-3}$, respectively, the total mismatch remains much closer to that of the dominant $(2,2)$ mode, which provides the leading contribution to the signal for this configuration.

\begin{figure}[t]
    \centering
    \includegraphics[width=0.5\textwidth]{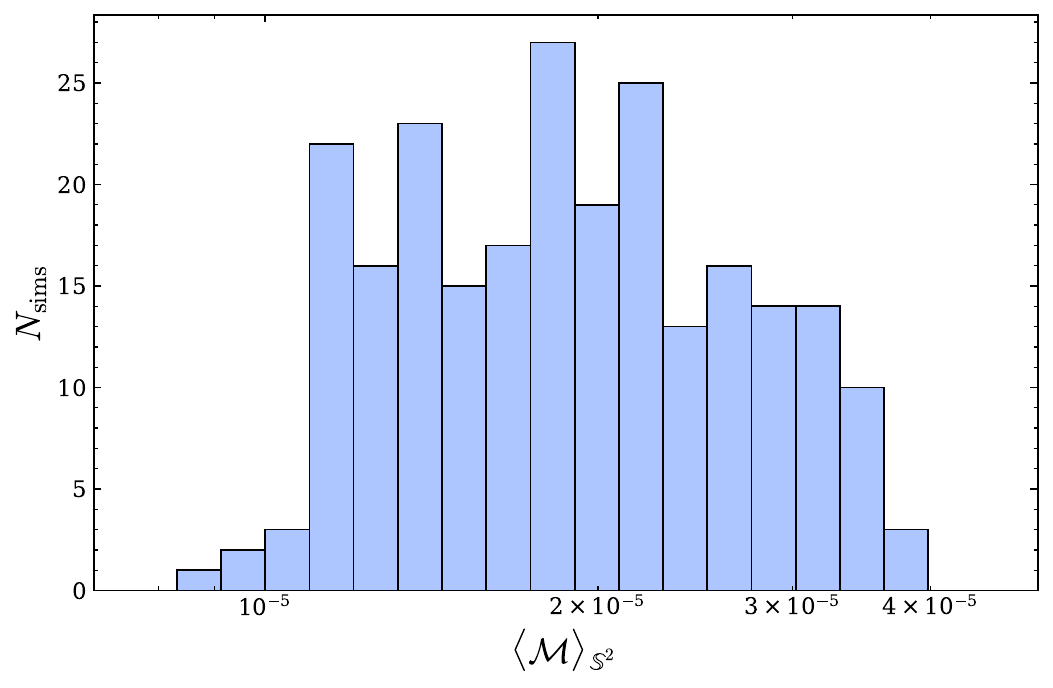}
    \caption{Distribution of the sky-averaged mismatch, $\langle\mathcal{M}\rangle_{\mathbb{S}^2}$, for the subset of quasicircular, nonprecessing SXS simulations considered in this work. The distribution is consistently concentrated around values of order $10^{-5}$.}
    \label{fig:mismatch_sphere_hist}
\end{figure}

We then extend the analysis to the subset of quasicircular, nonprecessing SXS simulations with ${\rm ID}>1000$ and $q<4$. For each simulation, we compute the mismatch map on the sphere and average it over all sky directions,
\begin{equation}
    \langle \mathcal{M} \rangle_{\mathbb{S}^2}
    =
    \frac{1}{4\pi}\int \mathcal{M}(\theta,\phi)\, d\Omega .
\end{equation}
The resulting distribution is shown in Fig.~\ref{fig:mismatch_sphere_hist}, and confirms that the sky-averaged mismatch remains consistently of order $10^{-5}$ across the sample.

%%%%%%%%%%%%%%%%%%%%%%%%%%%%%%%%%%
\section{\textsc{GreyRing} tests with synthetic data} \label{app:injrec}
%%%%%%%%%%%%%%%%%%%%%%%%%%%%%%%%%%
We validate our frequency-domain model through injection--recovery tests performed using the same waveform model for both the injection and the recovery. The signal includes the dominant multipole contributions $(\ell,\,m)=(2,\,\pm2)$ in the detector frame. In this frame, frequencies are redshifted by cosmological expansion, so that the measured mass corresponds to the redshifted detector-frame mass.

The intrinsic parameters of the injection are calibrated using the numerical relativity simulation \texttt{SXS:BBH:3617}, rescaled to a redshifted remnant mass $(1+z)M = 68.4\,M_\odot$, consistent with GW250114-like events. The corresponding dimensionless spin is $\chi \simeq \,0.686$. The phenomenological parameters $(A_{22},\,p_{22},\,c_{22})$ entering the amplitude and phase of the model (where we omit multipolar indices for ease of notation) are obtained by fitting the frequency-domain waveform of the simulation, following the procedure described in~\cite{Rosato:2025ulx}. The resulting values are reported in Table~\ref{tab:intrinsic}.

\begin{table}[h]
\centering
\caption{Intrinsic parameters used in the \textsc{GreyRing} injection.}
\begin{tabular}{c c c c c}
\hline
$(1+z)M/M_\odot$ & $\chi$ & $A_{22}$ & $p_{22}$ & $c_{22}$ \\
\hline
$68.4$ & $0.686$ & $5.588$  & $0.517$ & $1.191$ \\
\hline
\end{tabular}
\label{tab:intrinsic}
\end{table}

 We fix the extrinsic parameters (luminosity distance, sky location, inclination, polarization, reference time, and phase) to values consistent with those reported in Ref.~\cite{LIGOScientific:2025rid} for the ringdown analysis of GW250114. Indeed, for small-redshift sources, the luminosity distance $d_L$ is degenerate with the overall postmerger amplitude, and can therefore be neglected without loss of generality. Also, as we are only performing analyses with $\ell=|m|=2$ modes, the pattern functions will be the same, so we further fix the inclination angle. Finally, the sky location, reference time and phase are fixed, as their impact on the results is mild. This choice allows us to assess the performance of the model under realistic observational conditions. The adopted values are listed in Table~\ref{tab:extrinsic}.

\begin{table}[h]
\centering
\caption{Extrinsic parameters used in the injection.}
\begin{tabular}{c c c c c c c}
\hline
$d_L/{\rm Mpc}$ & $\theta$ & $\psi$ & RA & Dec & $t_{\rm geocent}/{\rm s}$  \\
\hline
$403.0$&$0.78$ & $1.329$ &$2.333$ &$0.190$ &$1420878141.235932$ \\
\hline
\end{tabular}
\label{tab:extrinsic}
\end{table}

\begin{figure*}
    \centering
    \includegraphics[width=0.9\linewidth]{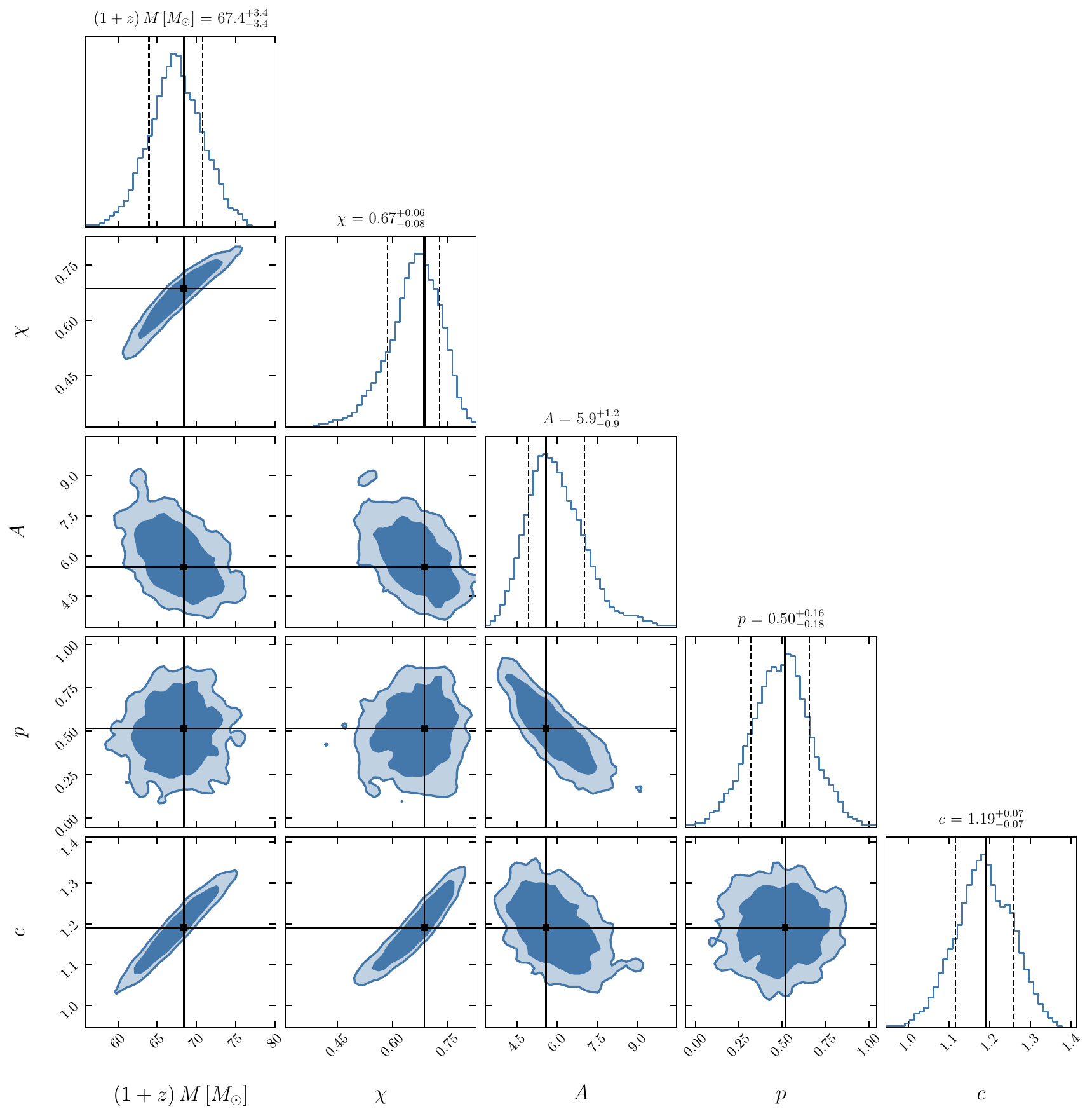}
    \caption{Joint posterior distributions for $((1+z)M,\, \chi,\, A_{22},\, p_{22},\, c_{22})$ from the injection--recovery analysis with \textsc{GreyRing}. Injected values are shown as solid black lines. Dashed lines in the one-dimensional panels mark the $68\%$ credible intervals, with medians reported above each panel. The darker and lighter blue regions in the two-dimensional panels correspond to the $68\%$ and $90\%$ joint credible regions, respectively. All injected parameters are accurately recovered, confirming the robustness of the model in the selected frequency range. Details of the injection are in the text.}
    \label{fig:injection}
\end{figure*}

\begin{figure*}[t]
    \centering    \includegraphics[width=0.9\linewidth]{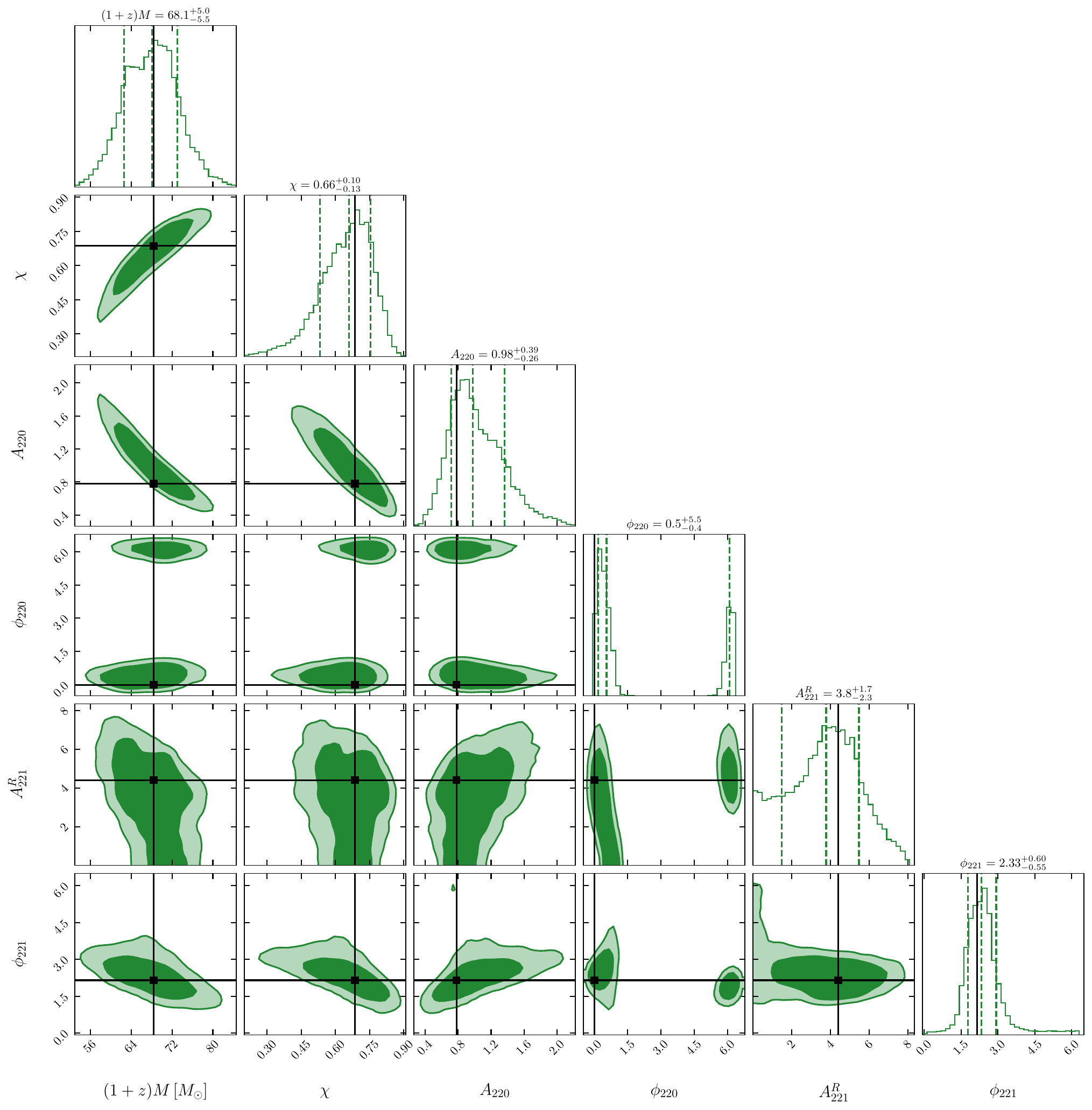}
    \caption{
    Joint posterior distributions for the parameters $((1+z)M,\, \chi,\, A_{220},\, \phi_{220},\, A_{221}^R,\, \phi_{221})$ obtained from the injection-recovery analysis using QNM-based BH spectroscopy with \texttt{pycbc inference}. The injected values are indicated by solid black lines. In the one-dimensional marginal panels, the dashed black lines mark the bounds of the $68\%$ credible intervals. In the two-dimensional panels, the darker and lighter regions denote the $68\%$ and $90\%$ joint highest-posterior-density credible regions, respectively, highlighting correlations among parameters. Details of the injection are given in the text.}
    \label{fig:injectionQNM}
\end{figure*}

The signal is injected into a network of ground-based detectors, namely LIGO Hanford and LIGO Livingston, assuming zero-noise data and adopting the Advanced LIGO design sensitivity~\cite{LIGO_T2200043_v3}. The resulting signal-to-noise ratio (SNR) is $24.12$ in Hanford and $27.38$ in Livingston, giving a total ${\rm SNR}=36.49$. The analysis is restricted to the frequency band $f \in [140,500]\,{\rm Hz}$.
The estimation of the parameters $(M, \chi, A_{22}, p_{22}, c_{22})$ is performed using \texttt{Bilby}~\cite{bilby_paper} with the \texttt{dynesty}~\cite{Speagle:2019ivv} sampler.

All injected parameters are accurately recovered within the $90\%$ credible intervals. The posterior distributions, shown in Fig.~\ref{fig:injection}, are sharply peaked around the injected values, indicating that parameter degeneracies are well controlled within the relevant frequency range. 

We have verified that the results are robust against variations of both the prior ranges and the frequency band. In particular, the recovery remains stable under changes of the lower and upper cutoffs, $f_{\rm min}$ and $f_{\rm max}$, within a reasonable range around their fiducial values.
We have also tested the relative-binning likelihood and found results to be consistent with those obtained using the standard likelihood. All results presented here are obtained with the standard likelihood. 

Finally, we note that there is a computational advantage in using \textsc{GreyRing} over traditional QNM analyses. Because Eq.~\eqref{eq:SMfullmodel} is explicitly a frequency-domain model, the likelihood can be evaluated with the typical $\mathcal{O}(N)$ complexity of frequency-domain analyses, in contrast with the higher computational cost of direct time-domain or gated-and-inpainted analyses (see Sec.~6 of Ref.~\cite{Berti:2025hly} for a detailed discussion).

\subsection{Injection-recovery with QNMs: comparison of graybody factors and QNMs}
\label{inj_rec_qnm}

Besides the analysis performed above, 
we perform an injection-recovery analysis with GW250114-like mock data, but described with a QNM model. Specifically, we inject a QNM-like signal with the mass and spin values reported in Table~\ref{tab:intrinsic}. The injected QNM signal contains the fundamental $(\ell,\,m,\,n)=(2,\,2,\,0)$ mode and the first overtone $(2,\,2,\,1)$, with amplitudes and phase values evaluated by using the fits of Ref.~\cite{Cheung:2023vki}, choosing as progenitor parameters those of \texttt{SXS:BBH:3617}: see Table~\ref{tab:intrinsic2}. As in the \textsc{GreyRing} analysis, the extrinsic parameters are fixed to the values reported in Table~\ref{tab:extrinsic}, and we use the same network and sensitivity curves. We perform the analysis with the \texttt{pycbc inference}~\cite{Biwer:2018osg} pipeline, using a gated-and-inpainted Gaussian likelihood noise model~\cite{Usman:2015kfa, Zackay:2019kkv, Capano:2021etf} to remove the influence of pre-peak times not associated with the ringdown. The strain data within a time interval $t\in[t_c+t_{\rm offset}-0.5\,{\rm s},\,t_c+t_{\rm offset}]$ are replaced/inpainted such that the filtered inverse power spectral density is zero
at all times corresponding to the chosen interval~\cite{Zackay:2019kkv}. In our case, we fix $t_{\rm offset}=10M\approx3.4\,{\rm ms}$. We report the estimated posterior parameters in Fig.~\ref{fig:injectionQNM}.

\begin{table}[h]
\centering
\caption{Intrinsic parameters used in the QNM injection, where we defined $A^R_{221}=\frac{A_{221}}{A_{220}}$.}
\begin{tabular}{c c c c c c}
\hline
$(1+z)M/M_\odot$ & $\chi$ & $A_{220}/10^{-20}$ & $\phi_{220}$ & $A^R_{221}$ & $\phi_{221}$ \\
\hline
$68.4$ & $0.686$ & $0.796$  & $0$ & $4.353$ & $2.180$ \\
\hline
\end{tabular}
\label{tab:intrinsic2}
\end{table}

\begin{figure}[t]
    \centering
    \includegraphics[width=\linewidth]{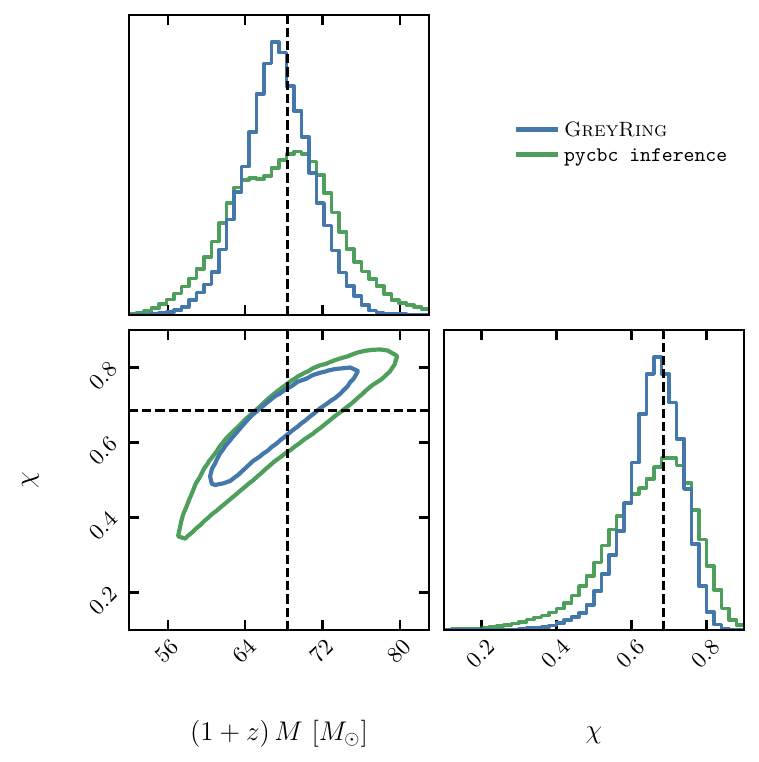}
    \caption{Comparison of the joint posterior distributions for $(M,\,\chi)$ obtained from injection--recovery analyses using the standard QNM-based model and \textsc{GreyRing}. The injected values are shown as dashed black lines. The contours correspond to the $90\%$ joint credible regions. While both models accurately recover the injected parameters, the \textsc{GreyRing} posterior appears slightly tighter.}
    \label{fig:cornercbcgreyring}
\end{figure}

To further assess the performance of the model, we directly compare the recovery of the remnant properties $(M,\,\chi)$ obtained with standard QNM-based spectroscopy and with \textsc{GreyRing}. This comparison isolates the determination of the fundamental physical parameters, independently of the specific modeling of the amplitude and phase.
In Fig.~\ref{fig:cornercbcgreyring} we show the corresponding corner plots for the two analyses. In both cases, the posterior distributions are centered around the injected values, indicating that both approaches provide an unbiased recovery of the remnant mass and spin.
The posterior obtained with \textsc{GreyRing} is slightly tighter than the one derived from the QNM model. 
Overall, the two approaches yield consistent results and similar $M-\chi$ correlations, with \textsc{GreyRing} offering a modest improvement in parameter estimation within the considered frequency range.

%%%%%%%%%%%%%%%%%%%%%%%%%%%%%%%%%%
\section{\textsc{GreyRing} tests with real data} 
\label{app:realdata}
%%%%%%%%%%%%%%%%%%%%%%%%%%%%%%%%%%

We analyze publicly available strain data from the LIGO detectors using the \texttt{Bilby} framework~\cite{bilby_paper}. The data are accessed through the \texttt{GWpy} interface~\cite{GWpy} and consist of time-domain strain sampled at $4096\,\mathrm{Hz}$.

The waveform model discussed in Sec.~\ref{app:fittingprocedure} is defined in the frequency domain and provides the two polarizations $(h_+, h_\times)$, which are projected onto each detector using the standard antenna response functions. The extrinsic parameters (sky location, inclination, polarization, and distance) are fixed to the values reported in Table~\ref{tab:extrinsic}, while the reference time and phase are allowed to vary and analytically marginalized over.

The intrinsic parameters of the model, $(M,\, \chi,\, A,\, p,\, c)$, are sampled in the inference. Parameter estimation is performed using the \texttt{dynesty} nested sampler~\cite{Speagle:2019ivv} with $n_{\rm live}=1000$ live points. We have verified that the results are robust against variations of the sampler settings.

The analysis is restricted to a frequency band $[f_{\min},\, f_{\max}]$ with $f_{\max}=512\,\mathrm{Hz}$, where the data are noise-dominated. The lower cutoff $f_{\min}$ is varied within the range $40$--$200\,\mathrm{Hz}$ in order to probe the stability of the inference and to identify the regime of validity of the model.

\begin{figure}[t]
    \centering
    \includegraphics[width=\columnwidth]{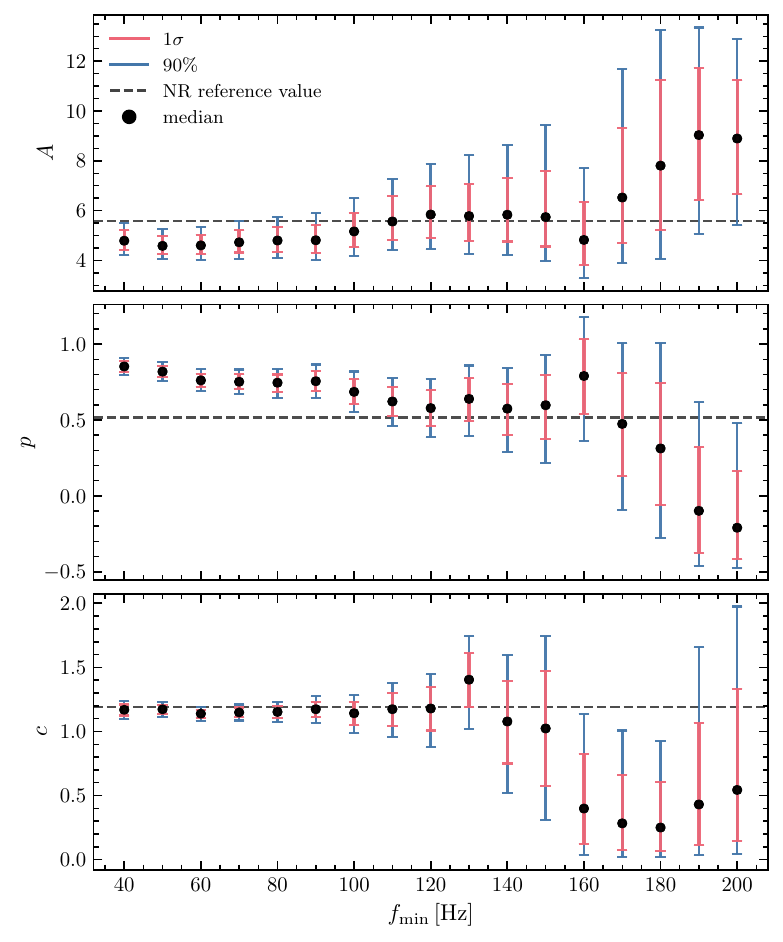}
    \caption{
    Posterior summary for the parameters $(A_{\ell m},\,p_{\ell m},\,c_{\ell m})$ as a function of the lower cutoff frequency $f_{\min}$, with $f_{\max}=512\,\mathrm{Hz}$. Black dots denote posterior medians, while red and blue bars indicate $1\sigma$ and $90\%$ credible intervals, respectively. The dashed horizontal lines mark the reference values obtained from the numerical relativity simulation \texttt{SXS:BBH:3617}, chosen to match the properties of the observed event. The results highlight an intermediate region, $f_{\min}\sim 100$--$140\,\mathrm{Hz}$, where the parameters are consistently recovered, while lower and higher frequency cutoffs lead to biased or weakly constrained estimates.  %\eb{Can you add "22" to $A_{22}$, $p_{22}$ and $c_{22}$ on the y-axes, please?}
    }
    \label{fig:params_fmin}
\end{figure}

In Fig.~\ref{fig:params_fmin} we show the inferred values of the phenomenological parameters $(A_{\ell m},\,p_{\ell m},\,c_{\ell m})$ as functions of the lower cutoff frequency $f_{\min}$.
The error bars of $A$ and $p$ in the stable frequency region are consistent with the expectations from a previous hyperfit analysis~\cite{Rosato:2025ulx}, where these parameters were modeled as functions of the properties of the progenitor binary. In particular, using the source parameters reported by the LVK Collaboration, the predicted values of $A$ and $p$ are in agreement with those inferred here.

\begin{figure*}[ht]
\centering
\includegraphics[width=0.66\columnwidth]{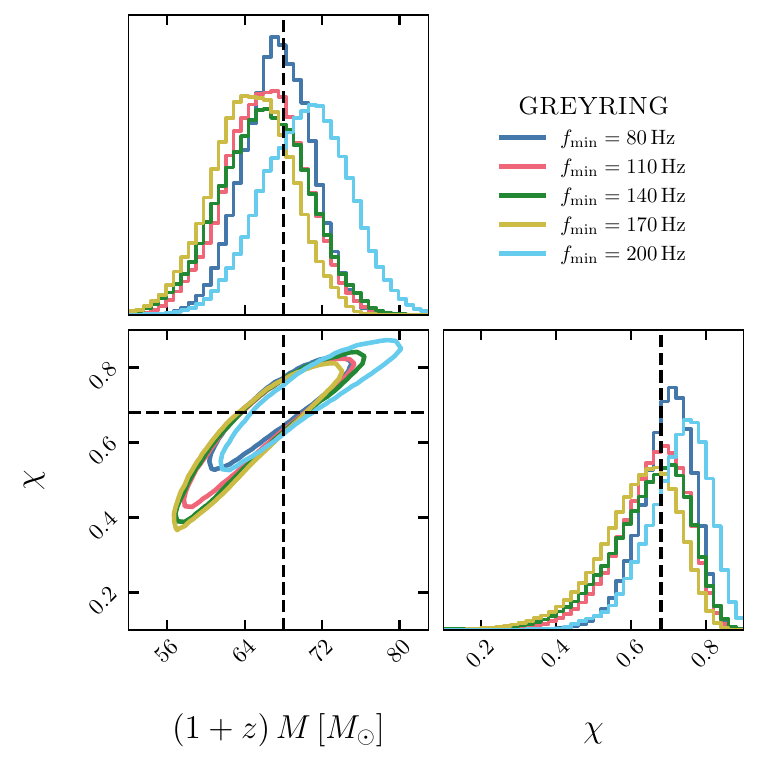}
\hfill
\includegraphics[width=0.66\columnwidth]{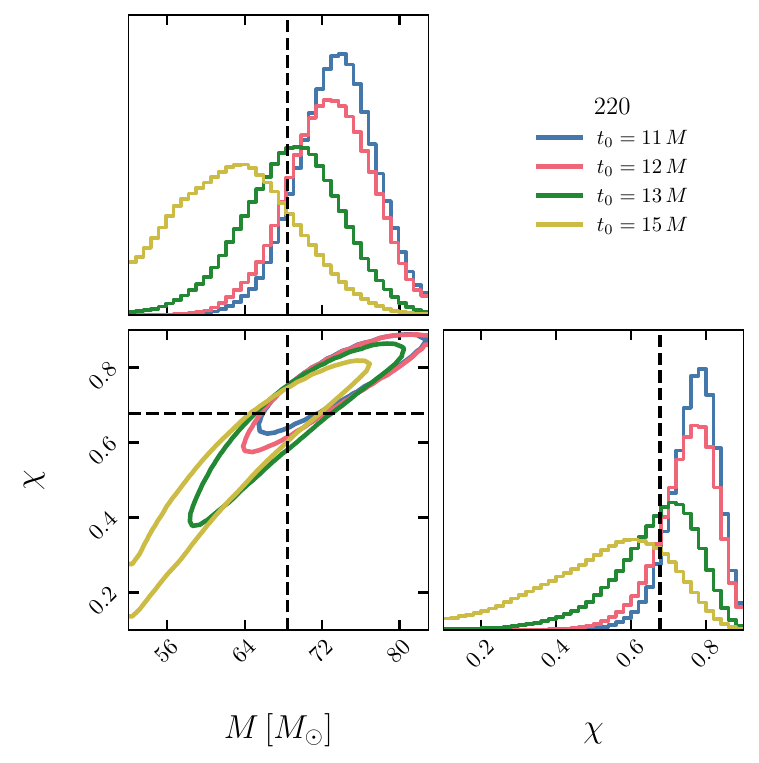}
\hfill
\includegraphics[width=0.66\columnwidth]{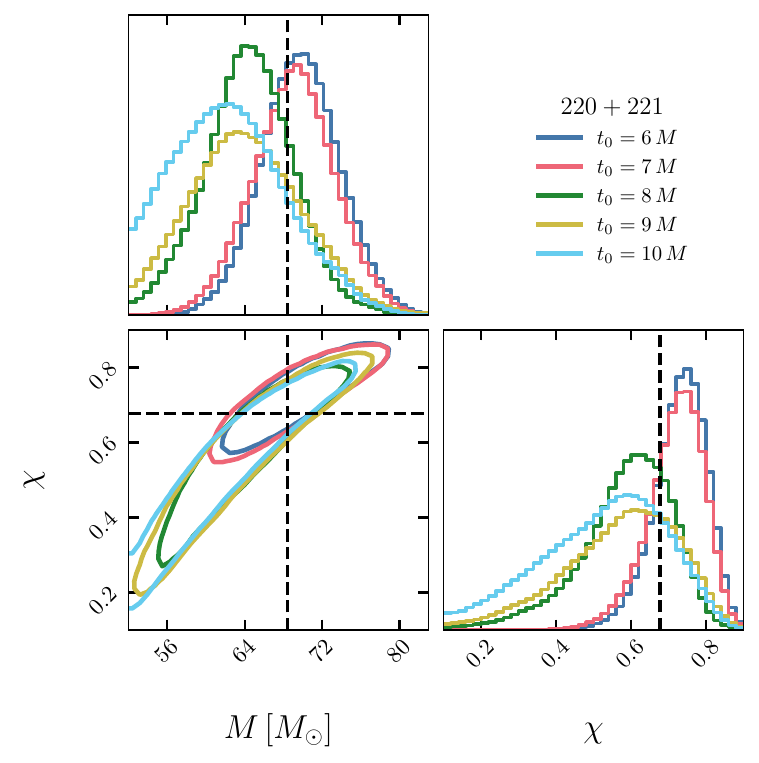}

\caption{
Left panel: \textsc{GreyRing} posterior distributions for the remnant mass $M$ and spin $\chi$ of GW250114, obtained for different lower cutoff frequencies $f_{\min}$ at fixed $f_{\max}=512\,\mathrm{Hz}$. Central and right panels: LVK QNM-based reconstructions of the same remnant parameters obtained by varying the ringdown start time $t_{\rm start}$, using either the fundamental mode alone, $(2,2,0)$, or the fundamental mode together with the first overtone, $(2,2,0)+(2,2,1)$, respectively. Contours enclose $90\%$ credible regions, and the corresponding one-dimensional marginalized distributions are shown along the diagonal. For clarity, only representative subsets of frequency cutoffs and starting times are displayed. The \textsc{GreyRing} posteriors remain stable over a broad range of $f_{\min}$ and are consistently compatible with the remnant parameters inferred from the LVK inspiral-merger-ringdown analysis. By contrast, the QNM-based posteriors exhibit a substantially stronger dependence on $t_{\rm start}$, both for the fundamental-mode-only analysis and when the first overtone is included. This comparison illustrates the robustness of the \textsc{GreyRing} remnant inference against variations of the frequency-domain analysis window, relative to the starting-time sensitivity of standard QNM-based spectroscopy.
}
\label{fig:Mchi_fmin}
\end{figure*}

At low values of $f_{\min}$, the recovered parameters are not stable and exhibit a clear bias with respect to the values recovered for intermediate values of  $f_{\min}$. This is expected, as in this frequency range the signal contains significant inspiral contributions that our model does not capture. Despite the small statistical uncertainties in this regime, driven by the larger SNR, the inferred parameters are systematically biased.

At high values of $f_{\min}$, the inference also deteriorates. In this case, the available signal is progressively reduced and the data become increasingly dominated by detector noise, leading to larger uncertainties and loss of constraining power.

An intermediate frequency range, $f_{\min}\sim 100$--$140\,\mathrm{Hz}$, provides a stable recovery of all parameters, with credible posterior intervals consistent with the reference values at the $90\%$ level in the region. This identifies the regime of validity of the model for the considered dataset.

This stability criterion is precisely what determines the regime of validity of the model for GW250114. The underlying logic is the same as in the analysis of numerical relativity simulations, where the validity range is identified by requiring the recovered parameters to stabilize as the lower cutoff is varied. For real data, we generalize this criterion to the full posterior distributions and select the frequency interval over which the inferred parameters remain mutually consistent, compatible with the reference values within their credible regions, and not yet significantly degraded by the loss of information caused by removing an excessive portion of the signal from the analysis band.

While the phenomenological parameters $(A_{\ell m},\,p_{\ell m},\,c_{\ell m})$ exhibit a nonnegligible dependence on the choice of the lower cutoff frequency, a different behavior is observed for the physical parameters of the remnant. In Fig.~\ref{fig:Mchi_fmin} we show the $90\%$ posterior contours for the remnant mass $M$ and spin $\chi$ obtained for representative values of the lower cutoff frequency $f_{\min}$.
In contrast to the phenomenological parameters, the inferred values of $(M,\,\chi)$ remain remarkably stable across the explored frequency range. Although small shifts appear at very low and very high values of $f_{\min}$, the posteriors largely overlap and are consistent within their uncertainties, as well as with those inferred from the LVK inspiral-merger-ringdown analysis.
For comparison, the same figure also shows the LVK QNM-based posteriors obtained by varying the start time $t_{\mathrm{start}}$, both for the fundamental $(2,2,0)$ mode alone and for the $(2,2,0)+(2,2,1)$ model. These posteriors exhibit a substantially stronger dependence on $t_{\mathrm{start}}$, making the contrast with the frequency-domain stability of the \textsc{GreyRing} inference directly apparent. This behavior is consistent with the known start-time sensitivity of standard QNM analyses~\cite{LIGOScientific:2025epi,LIGOScientific:2025obp}, and with the results of our independent inference performed with \texttt{PyCBC}.
This robustness reflects the fact that $M$ and $\chi$ set the overall frequency scale and structure of the ringdown signal, which are already well constrained within a relatively narrow frequency interval. Consequently, their estimation is only weakly sensitive to the precise choice of the frequency band and to residual modeling inaccuracies at low frequencies.

Although the results shown in the plots refer to the inference in which all extrinsic parameters are fixed to the values reported in Table~\ref{tab:extrinsic}, we have also performed additional runs in which the inclination angle $\theta$ and the polarization angle $\psi$ are sampled rather than fixed, as a consistency check. We find that this has a negligible impact on the posterior distributions of the remnant mass and spin, confirming that our ability to constrain the remnant parameters is relatively insensitive to the treatment of these extrinsic parameters in the present setup.

\begin{figure*}[t]
    \centering
    \includegraphics[width=0.9\textwidth]{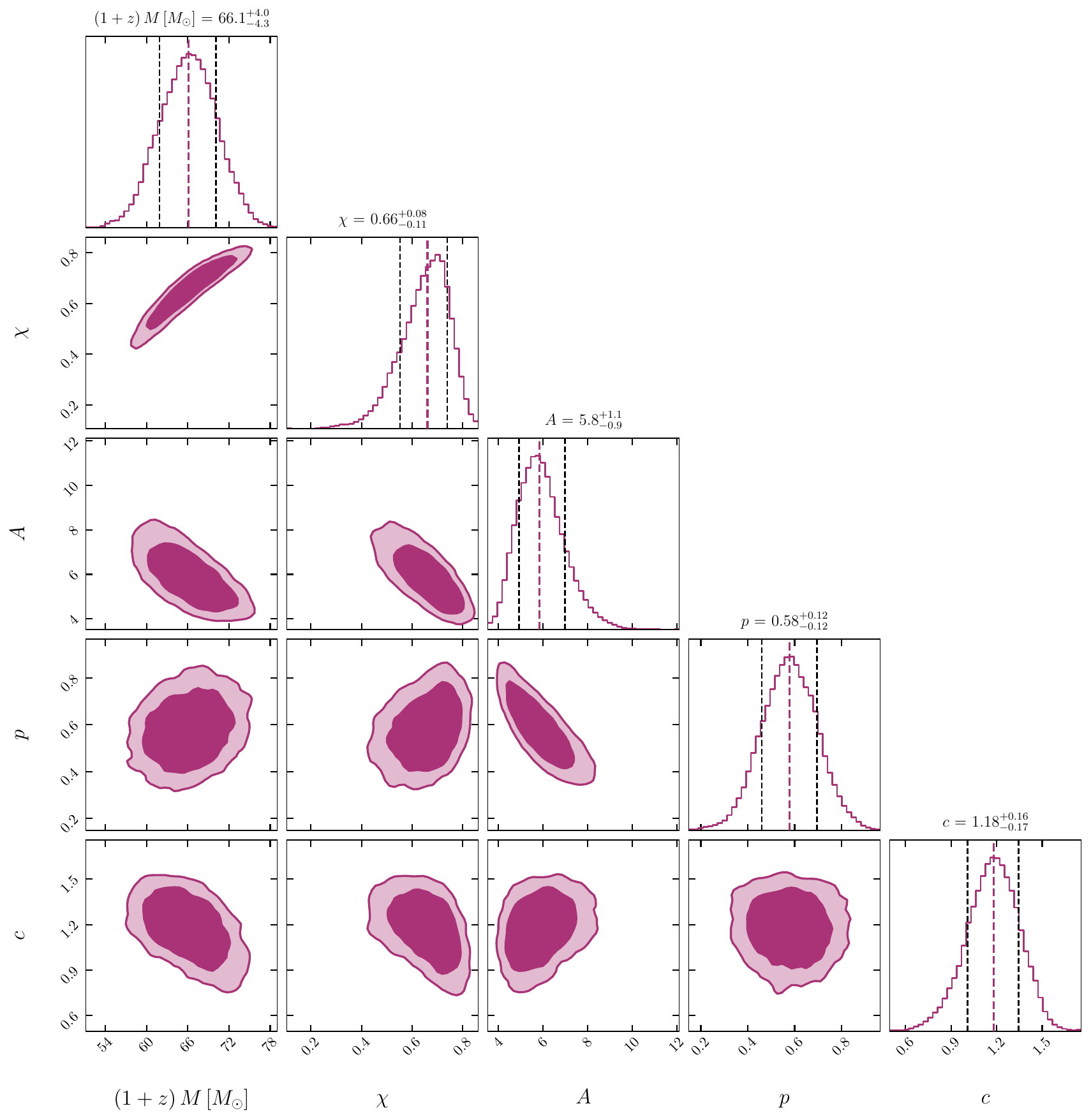}
    \caption{
    Posterior distributions and correlations for all model parameters, obtained for $f_{\min}=110\,\mathrm{Hz}$ and $f_{\max}=512\,\mathrm{Hz}$. Contours correspond to $90\%$ credible regions, while one-dimensional marginals are shown along the diagonal. The plot highlights significant correlations among the phenomenological parameters, particularly between $A_{22}$ and $p_{22}$, while $M$ and $\chi$ remain comparatively well constrained and less correlated. This choice of $f_{\min}$ lies within the regime of validity identified in Fig.~\ref{fig:params_fmin}.
    }
    \label{fig:corner_full}
\end{figure*}

This suggests that the remnant parameters are mainly controlled by the global structure of the signal and are therefore more robustly inferred than the auxiliary parameters $(A_{\ell m},\,p_{\ell m},\,c_{\ell m})$ in the present formulation of the model. Although the physical interpretation of $(A_{\ell m},\,p_{\ell m},\,c_{\ell m})$ is not yet fully understood and may evolve with further progress in the Green's function description of the ringdown, the stability of the inferred $(M,\,\chi)$ already indicates that incomplete knowledge of these quantities does not prevent robust access to the remnant properties. This may represent an advantage over standard QNM data analysis, in which initial conditions can affect the inference more directly.

To further characterize the structure of the inference,  in Fig.~\ref{fig:corner_full} we show the full posterior distribution for all parameters, obtained for a representative choice of $f_{\min}=110\,\mathrm{Hz}$ within the validity regime identified above.
The corner plot highlights correlations among the phenomenological parameters, and in particular a degeneracy between $A_{22}$ and $p_{22}$, reflecting the freedom to redistribute the overall amplitude and spectral slope within the model. By contrast, $A_{22}$ and $p_{22}$ show no significant correlation with $c$.
The posteriors for $(M,\,\chi)$ are well localized, while the phenomenological parameters $A_{22}$ and $p_{22}$ display broader and more correlated distributions. This confirms that the remnant properties are primarily determined by the global structure of the signal, while the phenomenological parameters encode subleading features and are more sensitive to the choice of frequency range and to modeling assumptions.

As an additional benchmark, we performed an independent parameter estimation using standard QNM-based spectroscopy within the \texttt{pycbc inference} framework, obtaining posterior distributions for $(M,\,\chi)$ consistent with those reported by the LIGO-Virgo-KAGRA Collaboration. The results obtained with \textsc{GreyRing} are in agreement with these independent QNM-based analyses.

\end{document}